\documentclass[]{pasj01}
\usepackage{graphicx}
\usepackage{subfigure}

\Received{}
\Accepted{}
 
\begin{document} 

\title{ 
AGN selection in the AKARI NEP deep field \\with the fuzzy SVM algorithm}

\author{Artem \textsc{Poliszczuk}\altaffilmark{1}%
\thanks{Address: National Centre for Nuclear Research, ul.Pasteura 7, room 523, 02-093 Warsaw, Poland}}
\altaffiltext{1}{National Centre for Nuclear Research, ul.Pasteura 7, 02-093 Warsaw, Poland}
\email{artem.poliszczuk@ncbj.gov.pl}

\author{Aleksandra \textsc{Solarz},\altaffilmark{1}}

\author{Agnieszka \textsc{Pollo}\altaffilmark{1,2}}
\altaffiltext{2}{Astronomical Observatory of the Jagiellonian University, ul.Orla 171, 30-244 Krakow, Poland}

\author{Maciej \textsc{Bilicki}\altaffilmark{3,4}}
\altaffiltext{3}{Center for Theoretical Physics, Polish Academy of Sciences, al. Lotnik\'ow 32/46, 02-668, Warsaw, Poland}
\altaffiltext{4}{Leiden Observatory, Leiden University, P.O.Box 9513, 2300RA Leiden,
The Netherlands}

\author{Tsutomu T. \textsc{Takeuchi}\altaffilmark{5}}
\altaffiltext{5}{Division of Particle and Astrophysical Science, Graduate School of Science, Nagoya University, Nagoya 464-8602, Japan}

\author{Hideo \textsc{Matsuhara}\altaffilmark{6}}
\altaffiltext{6}{Institute of Space and Astronautical Science, Japan Aerospace Exploration Agency, 3-1-1 Yoshinodai, Chuo-ku, Sagamihara, Kanagawa 252-5210, Japan}

\author{Tomotsugu \textsc{Goto}\altaffilmark{7}}
\altaffiltext{7}{National Tsing hua University, No. 101, Section 2, Kuang-Fu
Road, Hsinchu, Taiwan 30013.}

\author{Toshinobu \textsc{Takagi}\altaffilmark{8}}
\altaffiltext{8}{Japan Space Forum, 3-2-1, Kandasurugadai, Chiyoda-ku, Tokyo 101-0062 Japan}

\author{Takehiko \textsc{Wada}\altaffilmark{6}}

\author{Yoichi \textsc{Ohyama}\altaffilmark{9}}
\altaffiltext{9}{Academia Sinica Institute of Astronomy and Astrophysics, P
.O. Box 23-141, Taipei 10617, Taiwan}

\author{Hitoshi \textsc{Hanami}\altaffilmark{10}}
\altaffiltext{10}{Physics Section, Faculty of Humanities and Social Sciences
, Iwate University, 020-8550, Morioka, Japan}

\author{Takamitsu \textsc{Miyaji}\altaffilmark{11}}
\altaffiltext{11}{Instituto de Astronom\'ia sede Ensenada, Universidad Nacional Aut\'onoma de M\'exico,
    Ensenada, BC, 22860, Mexico}

\author{Nagisa \textsc{Oi}\altaffilmark{12}}
\altaffiltext{12}{Tokyo University of Science, 1-3 Kagurazaka, Shinjuku-ku,
Tokyo 162-8601, Japan}

\author{Matthew \textsc{Malkan}\altaffilmark{13}}
\altaffiltext{13}{Department of Physics and Astronomy, UCLA, Los Angeles, CA,
90095-1547, USA}

\author{Kazumi \textsc{Murata}\altaffilmark{14}}
\altaffiltext{14}{Hosei University, 3-7-2 Kajino-cho, Koganei, 184-1584 Tokyo, Japan}

\author{Helen \textsc{Kim}\altaffilmark{13}}

\author{Jorge D\'{i}az \textsc{Tello}\altaffilmark{11}}

\author{The NEP \textsc{Team}\altaffilmark{}}


\maketitle

\begin{abstract}
The aim of this work is to create a new catalog of reliable AGN candidates selected from the AKARI NEP-Deep field. Selection of the AGN candidates was done by applying a fuzzy SVM algorithm, which allows to incorporate measurement uncertainties into the classification process. The training dataset was based on the spectroscopic data available for selected objects in the NEP-Deep and NEP-Wide fields. The generalization sample was based on the AKARI NEP-Deep field data including objects without optical counterparts and making use of the infrared information only. A high quality catalog of previously unclassified 275 AGN candidates was prepared.
\end{abstract}

\section{Introduction}

In the modern astronomy rapid growth of volume of incoming data implies a necessity of applying new tools, adjusted to the big data requirements. In particular, data growth becomes a serious issue in the case of sky surveys. One of the most popular solutions to this problem is the usage of machine learning algorithms. The support vector machine (SVM) algorithm is one of the most popular and efficient among them. The classical version of the SVM~(\cite{vapnik95}) has been applied to numerous astrophysical tasks, e.g. to create catalogs of various celestial sources \citep{solarz12, malek13, kurcz16, krakowski16, marton16, solarz17, toth17} as well as to detect and classify detailed structures of interstellar medium in the local Universe~\citep{beaumont11}. However, the classical version is unable to incorporate measurement uncertainties into the classification process. This issue can be overcome by applying a modified version of the SVM algorithm - the fuzzy SVM~\citep{linwang02,linwang04} referred to as fSVM. In the present work, a catalog of active galactic nuclei (AGN) from AKARI NEP-Deep data was prepared. The selection of AGN candidates was performed by applying a binary classification task based on the fSVM algorithm (i.e., AGNs vs the rest of objects) to the infrared AKARI NEP-Deep source catalog, which was not based on the optical counterparts.

The AKARI satellite mission was launched in February 22, 2006, and carried out a series of infrared photometric surveys from the near-infrared (NIR) to the far-infrared (FIR) passbands. In particular, it performed a deep survey of the north ecliptic pole (NEP). Both the region of observation and the wavelength range of the survey (NIR and MIR) create a great opportunity for searching for distant and dusty AGNs. High galactic latitude reduces the star and galactic dust pollution while the wavelength range (MIR in particular) provides the most crucial information for identification of the type 2 AGNs~(\cite{assef18}). A strong underrepresentation of type 2 AGNs with respect to the predictions in the most of modern catalogs~(\cite{stern12,huang17}) makes searching for dusty AGNs an important step to better understanding of AGN properties, their evolution and environmental dependencies. The efficiency of mid-IR detection of type 2 AGNs is based on the fact that the dusty torus re-emits a significant amount of accretion disk light in the MIR part of the spectrum. A sensitivity  comparable to the MIR type 2 AGNs is shown by X-ray surveys, however, observation time needed to collect the same amount of data is significantly smaller in the case of IR surveys~(\cite{stern12}). All of the previously mentioned advantages of the MIR AGN search make the AKARI NEP-Deep data uniquely suited to study distant type 2 AGNs. 

This work is organized as follows. In Section 2 description of the data used for training and generalization is given. Section 3 is a brief overview of the SVM algorithm model. In Section 4 the algorithm training process and feature selection are described. Evaluation of the performance of different classifiers is presented in Section 5. In Section 6 properties of the final catalog are discussed. Section 7 contains a brief summary and discussion of the results.

\section{The data}

The AKARI NEP-Deep survey covers the 0.4 deg$^2$ area around the NEP~\citep{matsuhara06}. Observations were carried out by the Infra-red Camera~\citep{onaka07} in nine filters, which were centered at 2 $\mu$m ($N2$), 3 $\mu$m ($N3$), 4 $\mu$m ($N4$), 7 $\mu$m ($S7$), 9 $\mu$m ($S9W$), 11 $\mu$m ($S11$), 15 $\mu$m ($L15$), 18 $\mu$m ($L18W$) and 24 $\mu$m ($L24$). '\textit{W}' letter next to the name of some of the filers ($S9W$, $L18W$) stands for wider filters, which covered a part of the spectrum corresponding to the nearest filters: $S9W$ covered the wavelength range of $S7$ and $S11$, while $L18W$ covered the range of $L15$ and $L24$.

A source catalog used in the present work was constructed for the previous SVM-based classification of the NEP-Deep data by~\citet{solarz12}. It was prepared by applying source extraction software SExtractor~\citep{bertinarnouts96} to field images made in each of the filters separately, and limiting the data set to the objects with an existing measurement in the $N2$ passband. Thus, this original catalog is N2-selected. Properties of this catalog are listed in Table~\ref{tab:solarzcat}.

\begin{table}[h!]
  \tbl{Properties of the used NEP-Deep catalog.\footnotemark[$*$] }{%
  \begin{tabular}{lllcc}
      \hline
      Name &  $\lambda _{ref}$\footnotemark[$1$] & N$_{s}$\footnotemark[$2$] & mag\_min\footnotemark[$3$] & mag\_max\footnotemark[$4$]\\
      \hline
      N2 & 2.4 & 23 325 & 12.34 & 26.86 \\
      N3 & 3.2 & 19 544 & 12.44 & 25.35 \\
      N4 & 4.1 & 18 753 & 13.33 & 24.74 \\
      S7 & 7.0 & 6 513  & 11.59 & 23.06 \\
      S9W& 9.0 & 6 507  & 11.86 & 21.95 \\
      S11& 11.0 & 5810  & 11.95 & 21.79 \\
      L15& 15.0 & 5589  & 10.43 & 19.96 \\
      L18W&18.0 & 5696  & 9.57  & 22.91 \\
      L24 &24.0 & 2417  & 13.62 & 20.73 \\
      \hline
    \end{tabular}}\label{tab:solarzcat}
\begin{tabnote}
\footnotemark[$*$] Objects with existing N2 measurement. \\ 
\footnotemark[$1$] Reference wavelength.\\
\footnotemark[$2$] Number of sources.\\
\footnotemark[$3$] Minimal magnitude in a filter.\\
\footnotemark[$4$] Maximal magnitude in a filter.\\
\end{tabnote}
\end{table}

The decrease in the number of objects with the increasing wavelength is caused both by the diminishing sensitivity of the passbands and by the increase of the brightness of the Cat's Eye Nebula (NGC~6543) occupying a part of the field at longer wavelengths.

Because of the supervised type of the machine learning method used to search for AGNs, the training dataset with known labels (or classes) of each of the data points was needed. In the present work the identification of the training sample was based on the optical spectroscopic data. Because of the small number of the available spectroscopic data from the NEP-Deep field, the training sample was constructed by cross-matching the AKARI NEP-Wide data~\citep{kim12} with spectroscopic observations of this region~\citep{shim13} performed by MMT/Hectospec~\citep{fabricant05} and WYIN/Hydra~\citep{barden93} spectrographs. Primary targets for spectroscopic observations described by~\citet{shim13} were MIR sources with fluxes limited at S11 (S11 $<$ 18.5 mag) and L18 (L18 $<$ 17.9 mag) passbands. These limits gave approximately 50\% completeness of the NEP-Wide data in the corresponding filters. This MIR-selected sample was additionally limited by R-band cuts: 16 mag $<$ R $<$ 22.5 mag for Hectospec and 16 mag $<$ R $<$ 21 mag for Hydra. The bright end limit was imposed in order to avoid saturation and the faint end limit was introduced to select objects bright enough to obtain spectra of a good quality. Next, the sample for spectroscopic observations was randomly chosen from the primary targets set limited in R and MIR bands. Beside this main target selection, a smaller number of special targets was selected, among them high-redshift galaxy or AGN candidates. The AGN targets for spectroscopic observations were chosen by applying color cuts described in~\citet{lee07}: N2-N4 $>$ 0 and S7-S11 $>$ 0. In addition, the X-ray information to mark XAGN~\citep{krumpe15} was used.

\begin{figure*}[t]
\centering     
\subfigure[N2 flux distribution of particular classes from the labeled set and unlabeled NEP-Deep source catalog.]{\label{fig:wideXdeepN2}\includegraphics[width=80mm]{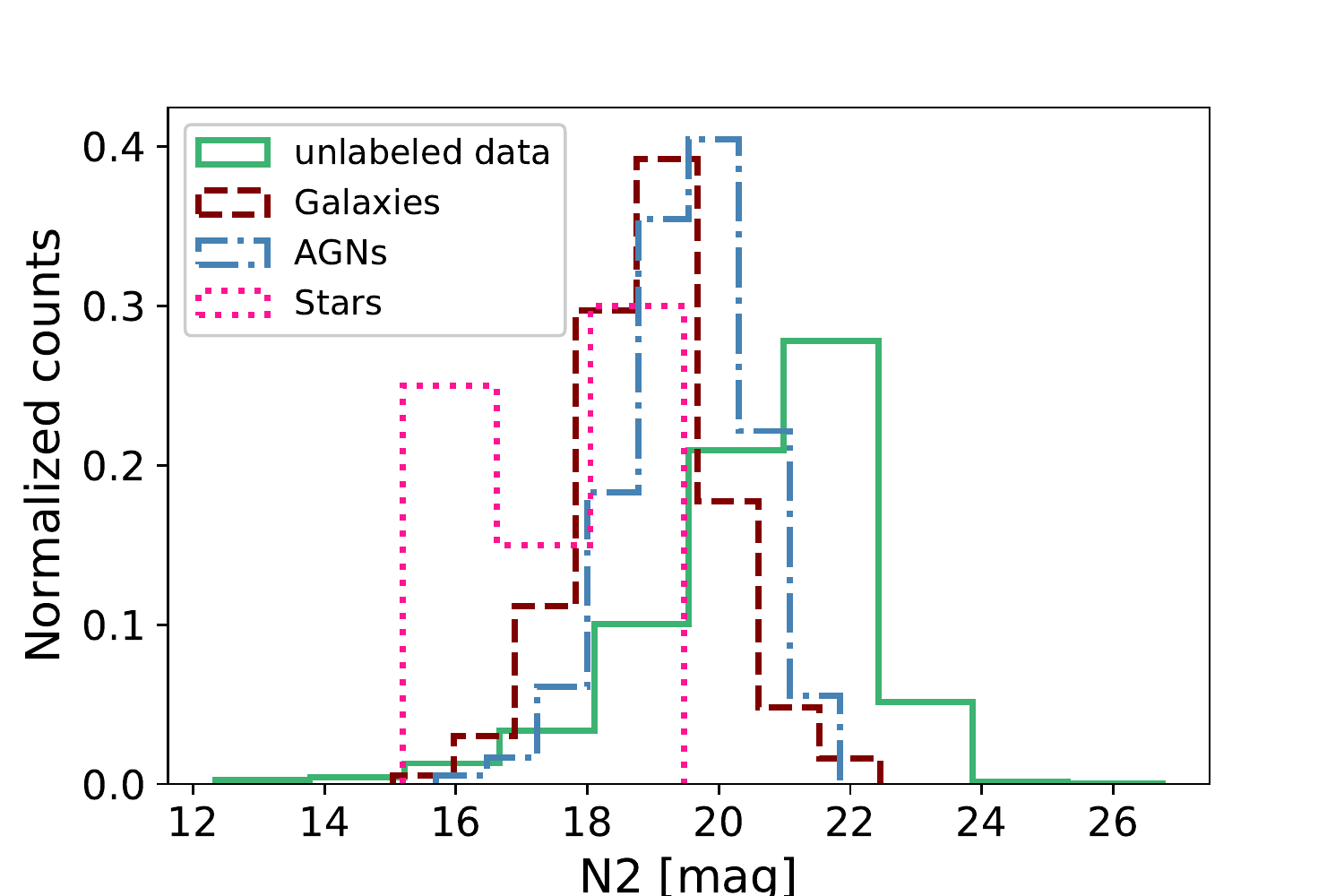}}
\quad
\subfigure[Redshift distribution of galaxies and AGNs with measured N2 flux.]{\label{fig:wideN2_Z}\includegraphics[width=80mm]{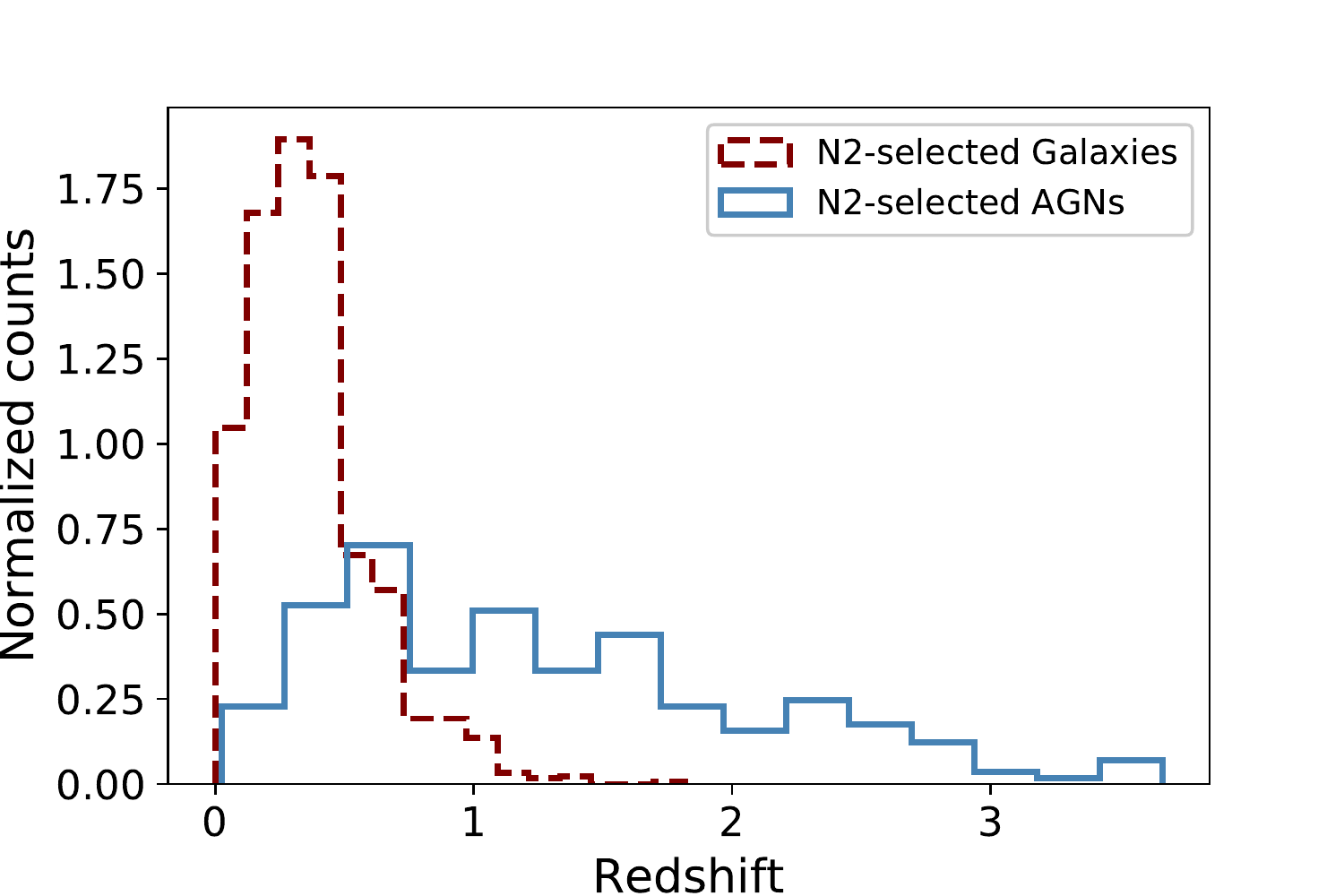}}
\caption{Selected properties of the generalization and labeled sets.}
\label{fig:n2_properties}
\end{figure*}

This sample based on~\citet{shim13} was enriched by additional small sample of spectroscopic NEP data gathered by Keck/DEIMOS~\citep{faber03}, GTC/OSIRIS~\citep{cepa00} and Subaru/FMOS~\citep{kimura10} telescopes and available in the internal database of the AKARI NEP collaboration. Properties of the labeled sample are listed in Table~\ref{tab:wideXdeep}. Histograms of the N2 magnitude of labeled set and unlabeled NEP-Deep source catalog are shown in Figure~\ref{fig:wideXdeepN2}. Redshift distribution of labeled objects is shown in Figure~\ref{fig:wideN2_Z}. The training sample construction is described in more detail in Section~4.3.

\begin{table}[h!]
  \tbl{Properties of the labeled catalog prepared for the purpose of this work based on NEP-Wide and NEP-Deep data with available spectroscopic classification.}{%
  \begin{tabular}{lllccc}
      \hline
      Name & N$_{s}$\footnotemark[$1$] & z$_{med}$\footnotemark[$2$]& z$_{max}$\footnotemark[$3$] & mag\_min\footnotemark[$4$] & mag\_max\footnotemark[$5$]\\
      \hline
      N2 & 1930 & 0.36 & 3.66 &  15.03 & 22.72 \\
      N3 & 1938 & 0.36 & 3.66 &  15.36 & 22.64 \\
      N4 & 1955 & 0.37 & 3.66 &  15.84 & 22.92 \\
      S7 & 1310 & 0.30 & 3.68 &  15.19 & 21.23 \\
      S9W& 1654 & 0.35 & 3.70 &  14.92 & 20.70 \\
      S11& 1612 & 0.37 & 3.70 &  14.92 & 20.70 \\
      L15& 1373 & 0.39 & 3.70 &  14.43 & 19.96 \\
      L18W&1383 & 0.39 & 3.70 &  14.08 & 20.06\\
      \hline
    \end{tabular}}\label{tab:wideXdeep}
\begin{tabnote}
\footnotemark[$1$] Number of objects measured in a particular filter. \\ 
\footnotemark[$2$] Median redshift.\\
\footnotemark[$3$] The highest redshift.\\
\footnotemark[$4$] Maximal magnitude in a filter.\\
\footnotemark[$5$] Minimal magnitude in a filter.\\
\end{tabnote}
\end{table}

\section{The SVM algorithm}

The SVM is a supervised learning algorithm. This implies the usage of data with a priori known labels for the training process. As a result, it is possible to create a classifier, which can be used on a sample of objects with an unknown assignment -  this stage of algorithm performance will be referred to as \textit{generalization}.

Even having data labeled, it is still often very difficult or even impossible to separate classes in the input feature space (called \textit{input space}). The main idea of the SVM algorithm is to map the data into the high dimensional feature space (called \textit{feature space}), where the construction of separating hyperplane becomes possible. An output of such a classifier relies on the position of classified objects with respect to the separation hyperplane. It can be written as

\begin{equation}
f(\textbf{x}) = \sum^n_{i=1} \alpha_i k\left(\textbf{x},\textbf{x'}\right) + b,
\end{equation}

\noindent
where $k\left(\textbf{x},\textbf{x'}\right)$ is the kernel function, which in the SVM formulation can substitute the information about the real mapping from input space to feature space, $\alpha_i$ is a linear coefficient, and $b$ is bias, which refers to the perpendicular distance to the hyperplane.

In the classical SVM formulation there is no possibility to incorporate measurement uncertainties into the classification process. The fuzzy SVM allows to assign so called \textit{fuzzy memberships} to each of the training points. These memberships affect the importance of a particular training object in the hyperplane construction process. If fuzzy memberships are based on measurement uncertainties in a way which maximizes the importance of the training objects for the most precise measurement, one would get a more reliable and physically motivated classification.

In the present work fuzzy memberships were constructed as

\begin{equation}\label{eq:eFSVM_s}
		s_{i} = 1 - \frac{ err({\textbf{x}_{i}}) }{E^{max}_{\pm} + \delta}.
\end{equation}
An expression $err({\textbf{x}_{i}})$ in the numerator is defined as 
\begin{equation}
	err({\textbf{x}_{i}}) = \sum_{j=1}^{n} e_{ij} ,
\end{equation}
where $e_{ij}$ is a measurement uncertainty of the i-th 
object's j-th feature. A parameter $E^{max}_{\pm}$  is defined as a maximal 
$err({\textbf{x}_{i}})$ in a particular class $y$:
	\begin{equation}
			E^{max}_{\pm} = \max_{\textbf{x}_i : y = \pm 1} 				err({\textbf{x}_{i}}).
		\end{equation}
and $\delta$ is some small constant included to avoid the case $s_i$=0. In the 
current work $\delta$ was chosen to be equal to 10$^{-4}$.

A detailed discussion on fuzzy SVM and its comparison to the classical SVM algorithm will be presented in~\citet{poliszczukinprep}.

\section{Construction of the classifier}

In this section specification of tools used for training and evaluation is given. The code written for the present work was based on the scikit-learn library~\citep{scikit}, which incorporates lib-svm library~\citep{libsvm}. All classifiers presented here were binary classifiers trained on samples of AGN vs the rest of the sample (galaxies and stars).

\subsection{Training process and classifier specifications}

Training of the SVM classifier is based on searching for the optimal hyperplane for a given training data and tuning parameters of a model, including the topology of the separating hyperplane governed by the kernel function. In the present work two types of kernel functions were used. The radial basis kernel function (RBF)

\begin{equation}\label{eq:rbf}
	k\left(\textbf{x}_i,\textbf{x}_j\right) = \exp \left(- \gamma \parallel 					\textbf{x}_i-\textbf{x}_j \parallel^2 \right)
\end{equation}
and sigmoid kernel function
\begin{equation}\label{eq:sigm}
	k\left(\textbf{x}_i,\textbf{x}_j\right) = \tanh\left( \gamma 					\textbf{x}_i \cdot \textbf{x}_j + coef0 \right).
  \end{equation}
Another very popular kernel function - a polynomial kernel was not used in this work due to its high computational cost.

Parameters that can be tuned are the parameters of kernel functions. In the case of the RBF kernel it is $\gamma$, which defines how much of importance a single training point has. In the case of the sigmoid kernel a pair of parameters $\left(\gamma, coef0\right)$ is used. One of the advantages of the fuzzy SVM in comparison to the classical SVM is that fuzzy memberships can be treated as additional free parameters - this aspect of the fuzzy SVM makes it more flexible in the hyperplane construction. Another parameter that needs to be tuned is a regularization parameter (denoted by $C$), which controls the amount of misclassifications which can be made in a training process in order to obtain the best generalization ability. 

To obtain the best combination of parameters one has to perform so called \textit{grid search} process~\citep{burges98}, where different combinations of parameters are used to fit the training data and to obtain the most efficient classifier. The values of a parameter grid used in the present work are shown in Table~\ref{tab:grid}.

\begin{table}[h!]
  \tbl{Grid of the parameter values}{%
  \begin{tabular}{l|ccccccc}
  \hline
		C values: & 0.001 & 0.01 & 0.1 & 1 & 10 & 100 & 1000\\ 
		\hline 
		$\gamma$ values:& 0.001 & 0.01 & 0.1 & 1 & 10 & 100 & 1000 \\ 
		\hline 
        coef0 values: &0.001 & 0.01 & 0.1 & 1 & 10 & 100 & 1000 \\ 
\hline
\end{tabular}}\label{tab:grid}
\end{table}

After the grid search is finished, one has to search trough the volume of parameter space near the point of the best combination. This stage of the training is called \textit{deep} grid search. Parameters found in such a procedure were used in the present work for generalization on the unlabeled data. In the present work the 4-loop deep grid search was used, where each iteration used a grid with half-values of the parameters from the previous iteration.

\subsection{Evaluation metrics}

To find the best parameter combination in the grid search process one must be able to compare performance of classifiers with different parameter values. Such a performance measure is called the \textit{evaluation metric}~\citep{japkowicz14} and the grid search process is based on the maximization of one or multiple evaluation metrics. To be reliable, metric value is calculated on the set which was not used in the training process. Such a set will be referred to as a \textit{test set}.

Maximization of the metric value can carry the risk of leaking of the information about the distribution of the test set to the next grid search iterations. As a result, a classifier tries 
to fit to a test set and, as a consequence, its real generalization ability is overestimated by metrics. Such a situation is called \textit{overfitting}~\citep{bishop06}. 

One of methods of avoiding overfitting is a \textit{k-fold cross validation}~\citep{bishop06}, where instead of a simple training-test data division, a whole labeled dataset is divided into $k$ subsets. A final metric value is calculated in an iterative process: elements of $k-1$ subsets are used as a training set and elements of the last subset are used for testing and metric evaluation. As a consequence a set of $k$ different metric results is obtained. The final metric value is calculated by taking a mean value of $k$ results. In the present work 5-fold cross validation was used.

Metric used for the maximization in the grid search process was an area under the receiver operating characteristic curve (\textit{ROC curve},~\cite{fawcett06}), which will be referred to as \textit{ROC AUC}~\citep{bradley96}. The ROC curve describes the efficiency of a binary classifier with a change of the discrimination threshold and is created by plotting true positive rate (TPR) against false positive rate (FPR).

True positive rate or recall (which is a more common name in the case of its usage as a single metric) is defined as
\begin{equation}
\textrm{TPR} = \textrm{recall} = \frac{\textrm{TP}}{\textrm{TP}+\textrm{FN}}.
\end{equation}
In the case of a binary classifier one has to assign one of two class flags to an object. These flags can be referred to as positive or negative (in the case of this work positive class corresponds to the AGN class and negative - to the not-AGN or "Other" class). The TPR metric is defined as a ratio of the number of properly classified positive objects TP (True Positive) to the number of positive objects in the sample TP+FN (True Positive + False Negative). As a consequence one can interpret TPR (or recall) as a measurement of the completeness of the catalog of positive objects.

False positive rate (or the probability of the false alarm) is defined as 
\begin{equation}
\textrm{FPR} = \frac{\textrm{FP}}{\textrm{TN}+\textrm{FN}},
\end{equation}
or the ratio of the false positive candidates to all the objects classified as negative.

Both metrics, TPR and FPR, can have values from 0 to 1, where 1 corresponds to the highest value of the metric. In the simplest interpretation, the ROC curve describes a better classifier if the area under the curve is bigger. The random classification corresponds to the straight line connecting points (0,0) and (1,1). To generate the ROC curve for a single SVM classifier one has to calculate posterior probabilities for the classified objects. In the present work Platt's algorithm for the creation of the posterior probabilities was used~\citep{platt99}. An output of the SVM classifier produces an uncalibrated value, which represents the distance of the object from the decision surface. The Platt's method is based on fitting a parametric form of sigmoid function:
\begin{equation}
P\left(y=1|f\right) = \frac{1}{1+\exp(Af+B)},
\end{equation}
which maps the SVM output \textit{f} to posterior probability.

In addition to ROC AUC metric, which was maximized in the grid search process, other metrics were used for additional evaluation of the classifier performance. One of them, beside recall mentioned above, was precision, which is defined as 
\begin{equation}
\textrm{Precision} = \frac{\textrm{TP}}{\textrm{TP}+\textrm{FP}},
\end{equation}
or the ratio of the number of properly classified positive objects to all positive candidates. This formula can be interpreted as the measure of the purity of the output catalog of positive objects. Precision, just as recall or ROC AUC, can take values from 0 to 1.

One of the most crucial properties of the evaluation metric is its sensitivity to the size of the classes in the labeled sample. In the case of imbalanced data, which occurs in the present work (and is discussed in detail in the next subsection), a proper metric should be chosen for the correct evaluation of the performance on a smaller class. The ROC AUC is such a metric, because of its insensitivity to the imbalanced data. Maximization of the ROC AUC allows to optimize the completeness of the catalog without unnecessary loss of its purity~\citep{fawcett06}.

In the present work two additional popular metrics for unbalanced data application were used. One of them was Cohen's Kappa Coefficient (later referred to as kappa) which describes the agreement between real and predicted labels and can have values from 0 to 1, where 1 corresponds to the perfect agreement and 0 corresponds to the random classification~\citep{cohen60}. The second one is the Matthews Correlation Coefficient (later related to as matthews), which also describes a correlation between true and predicted labels and can have values from -1 to +1~\citep{matthews75}. The Matthews coefficient is well known to be slightly more sensitive to the performance of the classifier on the smaller class than Cohen's Kappa.

Both recall and precision are sensitive to the imbalanced data. Possible interpretation of their values as, correspondingly, completeness and purity of the catalog of positive candidates makes them important indicators of the classifier's efficiency.

\subsection{Feature selection}

A decreasing efficiency of the AKARI Infrared Camera at longer wavelengths and the resultant decrease of the number of objects leads to a number of input features dilemma. Bigger number of passbands gives more complete information about particular objects and simplifies the classification task. However, at the same time, it strongly reduces the number of objects and yields a more complicated selection function. A smaller number of filters gives bigger training and unlabeled sets but reduces the amount of information about data and makes the classification process more difficult. On the other hand, from the machine learning perspective, a bigger number of features can lead to overfitting. 

In the present work different feature selection strategies were tested. Physically motivated feature sets, and feature sets constructed using feature selection techniques were both used. In the case of feature selection methods, a relatively small training set available for AKARI NEP data was a strong limitation factor. Small amount of labeled data makes the proper evaluation of the performance of an algorithm a challenging task. Because of that, feature selection techniques based on application of machine learning algorithms to find the best feature set for a particular model, so called \textit{wrapper} techniques~\citep{kohavi97}, were not used in the present work. A \textit{filter} feature selection technique based on \textit{mutual information} (MI) was used instead~\citep{vergara14}. Filter methods assume complete independence between the learning algorithm and the data and therefore are much less prone to the risk of overfitting. 

Mutual information is a measure of statistical independence between two variables and is based on properties of the Shannon entropy. Defining the Shannon entropy of a variable \textit{x} as
\begin{equation}\label{eq:entropy}
H(x) = -\sum_{i=1}^{n} p(x_i)\log_2 \left(p(x_i)\right),
\end{equation}
which can be interpreted as a measure of its uncertainty, one can also define a conditional entropy
\begin{equation}\label{eq:conditionalentropy}
H(x|y) = \sum_{j=1}^{n} p(y_j)H(x|y=y_j),
\end{equation}
where the $H(x|y=y_j)$ is the entropy of all $x_i$, which are associated with $y=y_j$. The conditional entropy measures the remaining uncertainty of the random variable $x$ when the value of the random variable $y$ is known. The minimum value of the $H(x|y)$ appears when these two values are statistically dependent and there is no uncertainty in $x$ when the $y$ is known. The mutual information between $x$ and $y$ can be defined as
\begin{equation}\label{eq:mutualinfo}
MI(x;y) = H(x) - H(x|y),
\end{equation}
which is a non-negative quantity with the zero value corresponding to the statistically independent variables. The mutual information is widely used in the machine learning feature selection process because of its ability to capture any relation between two features, even if it is strongly nonlinear~\citep{vergara14}. The main weakness of this method is that it may fail to select the best feature set for a particular algorithm and data. The mutual information feature selection applied in this paper was based on the standard scikit-learn implementation.

In the present work six different feature sets were examined: 
\begin{enumerate}
\item A feature set containing only measurements in the NIR filters and colors corresponding to the neighboring passbands: N2, N3, N4, N2-N3, N3-N4 (5 features). This feature set will be referred to as $nir$.

\item A feature set containing only measurements in the MIR narrow filters (except L18W, which was used instead of L24) and colors corresponding to the neighboring passbands: S7, S11, L15, L18W, S7-S11, S11-L15, L15-L18W (7 features). This feature set will be referred to as $mir$.  

\item A feature set containing measurements of the NIR and wide MIR filters and corresponding colors of neighboring passbands: N2, N3, N4, S9W, L18W, N2-N3, N3-N4, N4-S9W, S9W-L18W (9 features). This feature set will be referred to as $wide\_mir$.  

\item A feature set constructed from the colors of neighboring NIR and narrow MIR passbands (except L18W, which was used instead of L24): N2-N3, N3-N4, N4-S7, S7-S11, S11-L15, L15-L18W (6 features). This feature set will be referred to as $6\_colors$.  

\item A set of 10 features constructed using mutual information feature selection method: N2, N2-N3, N2-N4, N2-S7, N2-S9, N2-L15, N2-L18, N3-N4, S11-L15, S11-L18. This feature set will be referred to as $mi\_features$.

\item A set of 10 features constructed using mutual information feature selection method on the second degree polynomial features obtained from the input feature set: N2-N4, N2(N2-N4), N3(N2-N4), N4(N2-N4), S7(N2-N4), S7(N3-N4), S9W(N2-N4), S11(N2-N4), L15(N2-N4), L18(N2-N4). This feature set will be referred to as $poly\_mi\_features$.

\end{enumerate}

The first two feature sets were constructed considering the strategy of preserving the higher number of unlabeled generalization data. Using the NIR and MIR passbands as separate feature sets allows us to work on the NIR source catalog without reducing its volume. Moreover, in the case of the high performance of classifiers trained on such separate feature sets, it would be possible to use the $mir$ classifier to select AGNs in the data without counterparts not only in optical but also in the NIR part of the spectrum. However, because of the importance of the MIR measurements in the process of AGN selection, such a separation makes the classifier ability to distinguish between the two classes much lower.

The $wide\_mir$ feature set was constructed as a compromise between the data volume preserving strategy presented in the $nir$ and $mir$ feature sets and the importance of the information contained in the MIR passbands. In addition to the NIR measurements, the data from the wide MIR passbands was also included. The wide MIR passbands do not define the MIR properties of the object as precisely as narrow passbands do, but they are available for a significantly bigger number of objects.

The $6\_colors$ feature set presents a widely applied physically motivated feature selection in the form of colors constructed from the neighboring filters. Its efficiency in the machine learning classification of the astronomical data was described in several publications~\citep{walker89, wolf01, solarz12}.

The MI-based feature sets can be treated as a reference point for physically motivated feature sets because of their purely statistical nature. Both $mi\_features$ and $poly\_mi\_features$ were selected as the most informative features during the MI feature selection from the set of all possible filters and colors (except for L24 filter) in the case of $mi\_features$ and from the set of all terms of the second degree polynomial constructed from all filters and all colors (except for L24 filter) in the case of $poly\_mi\_features$.

Uncertainties of the color values, which were used for the fuzzy membership construction were defined as
\begin{equation}\label{eq:color_uncertainty}
\sigma_{color} = \sqrt{\sigma_{f1}^2 + \sigma_{f2}^2},
\end{equation}
where $f1$ and $f2$ are values of fluxes used in the color definition and $\sigma_{f1}$, $\sigma_{f2}$ are their uncertainties. In the case of a feature constructed as a multiplication of color and flux, its uncertainty was defined as
\begin{equation}\label{eq:color_flux_uncertainty}
\sigma_{color \cdot flux} = \sqrt{(color \cdot \sigma_{flux})^2 + (flux \cdot \sigma_{color})^2}.
\end{equation}

Despite the reduction of the data volume based on the selection of different features, an additional limitation occurs on the stage of creation of the final catalog. Classical machine learning techniques can be considered reliable only in the region of the input parameter space occupied by the training sample. Moving outside this region leads to the extrapolation of the model to unknown distribution and can cause significant catalog contamination. This issue will be discussed in the next section.
Other selection effects that should be considered are: 
\begin{itemize}
\item classification model was based on the objects with optical counterparts which implies that the classifier is trained to search for similar types of objects,
\item all objects in the generalization set are detected in the N2 filter, i.e. we start from the N2-selected catalog.
\end{itemize}
Table~\ref{tab:trainsamples} shows numbers of objects in particular feature sets.

\begin{table*}
\tbl{Number of objects in different feature sets.}{%
\begin{tabular}{llllll}
\hline
Feature set & AGNs & Galaxies & Stars & Unlabeled & Number of features\\
\hline
nir                 & 227 & 1407 & 14 & 17 895 & 5 \\
mir                 & 189 & 686  & 6  & 1 994  & 7 \\
wide\_mir           & 193 & 889  & 7  & 3 109  & 9  \\
6\_colors           & 176 & 663  & 6  & 1 974  & 6   \\	
mi\_features        & 174 & 657  & 6  & 1 808  & 10   \\	
poly\_mi\_features  & 174 & 657  & 6  & 1 808  & 10   \\	
\hline
\end{tabular}}\label{tab:trainsamples}
\end{table*}

\section{Comparison of classifiers}

\subsection{Main comparison} \label{main_comparison}

In this Section results of the performance evaluation of different classifiers are shown. Because of different labeled sets used for training, the comparison between models based on different input feature space can be treated only as an approximation. However, a precise comparison between RBF and sigmoid kernels within the same parameter set can be made. The detailed results of the evaluation metrics are listed in Table~\ref{tab:evaluation}.

\begin{figure*}
\centering     
\subfigure[Evaluation metrics for all classifiers with the RBF kernel function. On the x axis different evaluation metrics are marked (ROC AUC score, Cohen's kappa, Matthews correlation coefficient, precision and recall), y axis shows values of these metrics obtained in the 5-fold cross validation with ROC AUC as maximized metric in the grid search process.]{\label{fig:rbfevalall}\includegraphics[width=80mm]{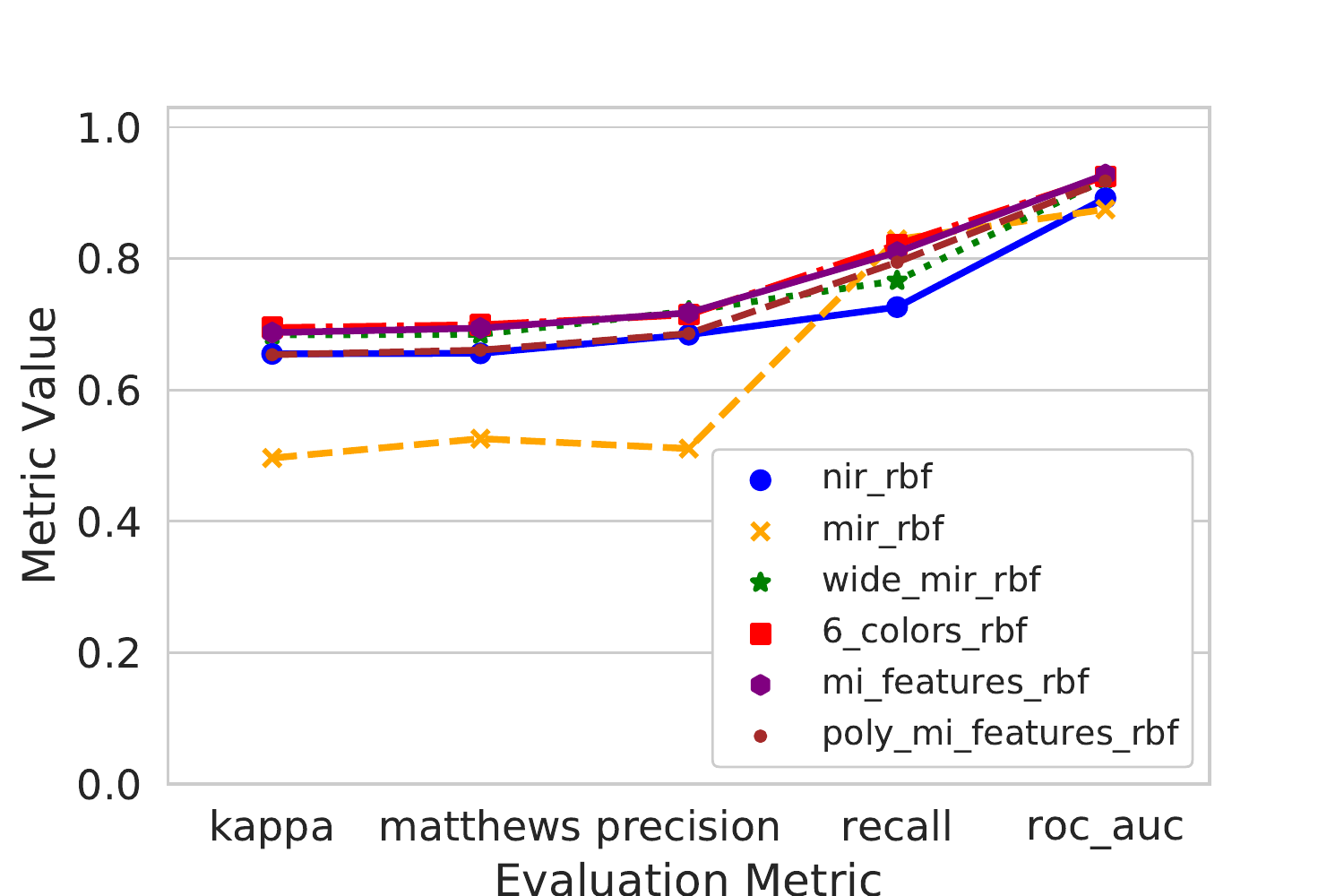}}
\quad
\subfigure[ROC curve for all classifiers with RBF kernel function. On the x and y axis false and true positive rate values, respectively, are shown. The curve was constructed from the mean values obtained in the 5-fold cross validation process on the best parameter combination]{\label{fig:rbfrocall}\includegraphics[width=80mm]{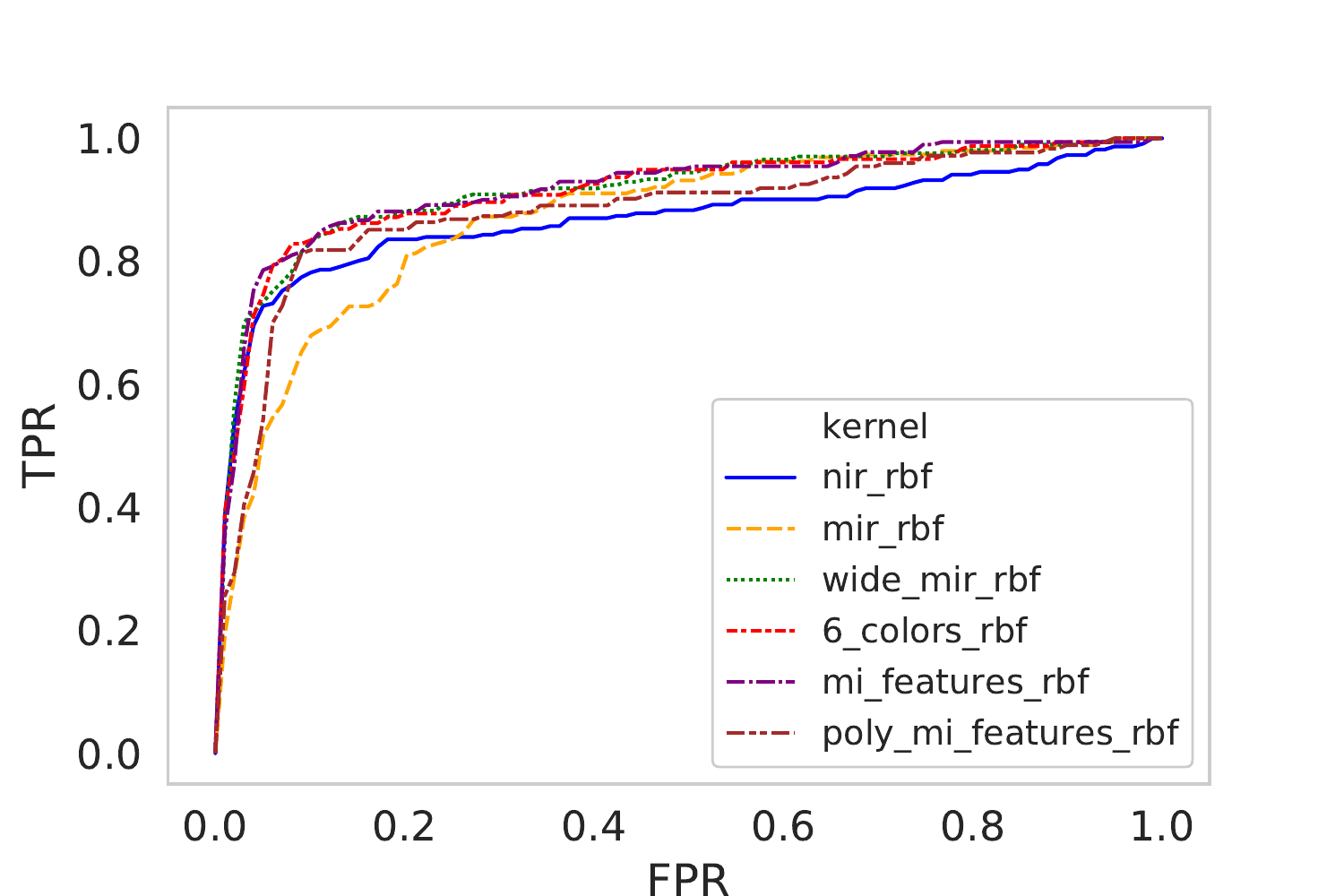}}
\caption{Performance of classifiers with RBF kernel trained on different feature sets.}
\label{fig:rbfall}
\end{figure*}

\begin{figure*}
\centering     
\subfigure[Evaluation metrics for all classifiers with the sigmoid kernel function. Explanation the same as in Fig.~\ref{fig:rbfevalall}.]{\label{fig:sigmevalall}\includegraphics[width=80mm]{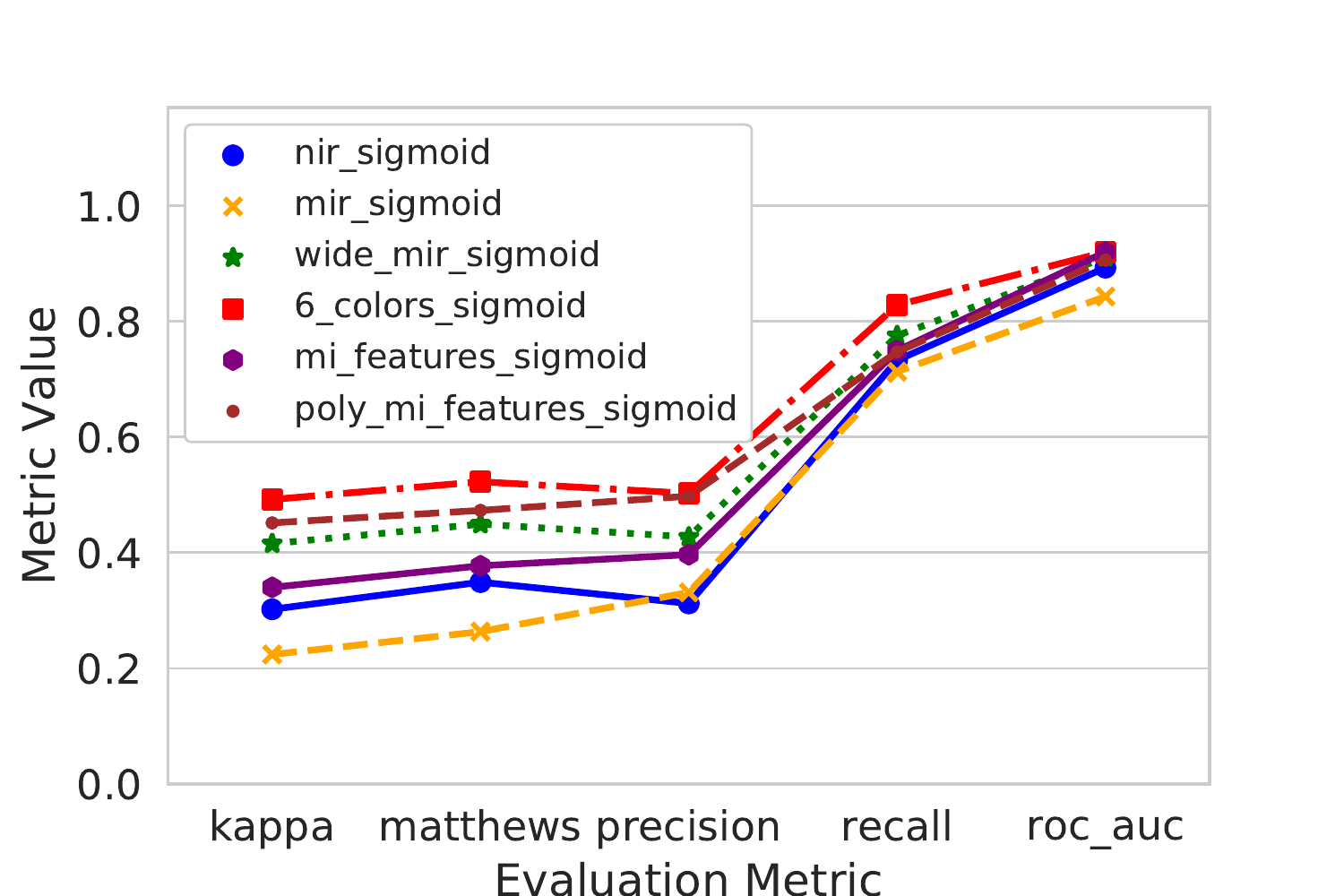}}
\quad
\subfigure[ROC curve for all classifiers with sigmoid kernel function. Explanation the same as in Fig.~\ref{fig:rbfrocall}.]{\label{fig:sigmrocall}\includegraphics[width=80mm]{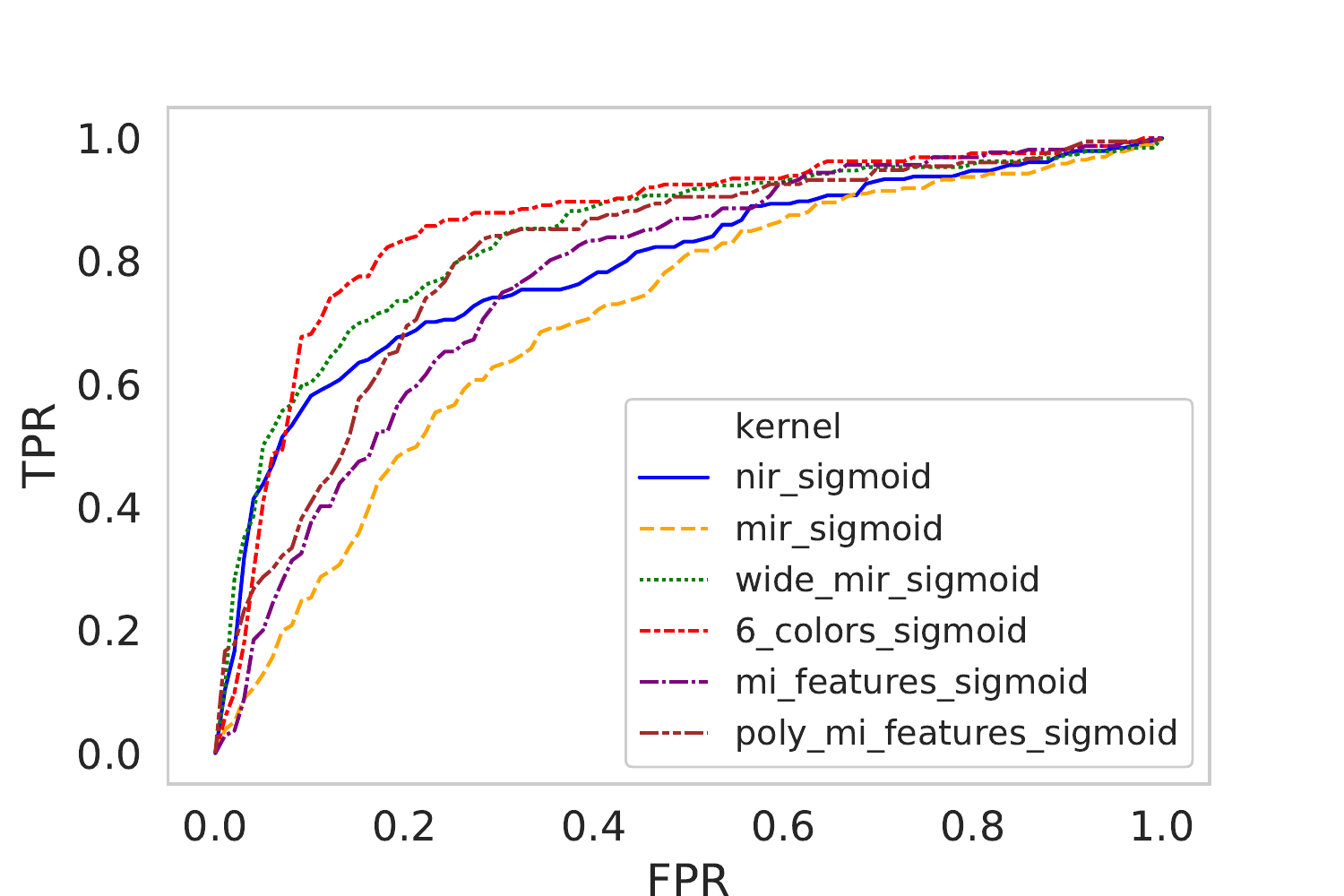}}
\caption{Performance of classifiers with sigmoid kernel trained on different feature sets.}
\label{fig:sigmoidall}
\end{figure*}

\begin{table*}
  \tbl{Evaluation of different classifiers. In columns 2 to 6 values of different evaluation metrics (ROC AUC, which was maximized in the grid search process, Cohen's kappa, Matthews correlation coefficient, precision and recall) with corresponding standard deviations are shown. Evaluation results and their uncertainties were obtained by performing 5-fold cross validation on the best parameter combination (obtained from the grid search).}{%
  \begin{tabular}{lccccc}
  \hline
	kernel    & roc auc         & kappa           & matthews        & precision       & recall \\
  \hline
nir\_rbf      & 0.89 $\pm$ 0.02 & 0.65 $\pm$ 0.03 & 0.66 $\pm$ 0.03 & 0.68 $\pm$ 0.04 & 0.73 $\pm$ 0.04\\
nir\_sigmoid  & 0.89 $\pm$ 0.02 & 0.30 $\pm$ 0.07 & 0.35 $\pm$ 0.07 & 0.31 $\pm$ 0.05 & 0.73 $\pm$ 0.08\\
mir\_rbf      & 0.87 $\pm$ 0.03 & 0.50 $\pm$ 0.06 & 0.53 $\pm$ 0.05 & 0.51 $\pm$ 0.06 & 0.83 $\pm$ 0.03\\
mir\_sigmoid  & 0.84 $\pm$ 0.04 & 0.22 $\pm$ 0.07 & 0.26 $\pm$ 0.08 & 0.33 $\pm$ 0.01 & 0.71 $\pm$ 0.1\\
wide\_mir\_rbf & 0.92 $\pm$ 0.01 & 0.68 $\pm$ 0.07 & 0.68 $\pm$ 0.07 & 0.72 $\pm$ 0.08 & 0.77 $\pm$ 0.04\\
wide\_mir\_sigmoid & 0.91 $\pm$ 0.01 & 0.42 $\pm$ 0.06 & 0.45 $\pm$ 0.05 & 0.43 $\pm$ 0.06 & 0.77 $\pm$ 0.04\\
6\_colors\_rbf     & 0.92 $\pm$ 0.03 & 0.69 $\pm$ 0.02 & 0.70 $\pm$ 0.02 & 0.71 $\pm$ 0.04 & 0.82 $\pm$ 0.06\\
6\_colors\_sigmoid & 0.92 $\pm$ 0.03 & 0.49 $\pm$ 0.05 & 0.52 $\pm$ 0.04 & 0.50 $\pm$ 0.06 & 0.83 $\pm$ 0.03\\
mi\_features\_rbf & 0.93 $\pm$ 0.04 & 0.69 $\pm$ 0.08 & 0.69 $\pm$ 0.08 & 0.72 $\pm$ 0.06 & 0.81 $\pm$ 0.13\\
mi\_features\_sigmoid & 0.92 $\pm$ 0.03 & 0.34 $\pm$ 0.06 & 0.38 $\pm$ 0.06 & 0.40 $\pm$ 0.07 & 0.75 $\pm$ 0.1\\
poly\_mi\_features\_rbf & 0.92 $\pm$ 0.04 & 0.65 $\pm$ 0.09 & 0.66 $\pm$ 0.09 & 0.69 $\pm$ 0.08 & 0.79 $\pm$ 0.1\\
poly\_mi\_features\_sigmoid & 0.91 $\pm$ 0.04 & 0.45 $\pm$ 0.13 & 0.47 $\pm$ 0.12 & 0.50 $\pm$ 0.12 & 0.75 $\pm$ 0.08\\
\hline
\end{tabular}}\label{tab:evaluation}
\end{table*} 
Visualizations of the metric values without standard deviations with additional ROC curves for the sigmoid and RBF kernels for all feature sets are shown in Figures~\ref{fig:rbfall} and~\ref{fig:sigmoidall}. All metric values exhibit prevalence of the RBF kernel over the sigmoid kernel for every feature set. Moreover, ROC curves show a stronger tendency to occupy the top left corner of the diagram in the case of RBF kernel - this tendency also demonstrates better performance and higher ROC AUC score of the classifiers with the RBF kernel. Based on these results, RBF kernel has been selected for the final classification. Consequently, only RBF kernel's performance will be discussed in the further analysis. From now on the "RBF" will be dropped from the description of the classifiers (it is left on the plots, though, to keep them self-explanatory).

The comparison between different feature sets is less intuitive because of different training samples. Visualizations of the evaluation metric values without standard deviations with additional ROC curves for all feature sets are shown in Figure~\ref{fig:rbfall}. In all cases the recall value is higher than precision because of indirect recall optimization in the process of ROC AUC maximization in grid search. Due to the similarity between kappa and matthews their behavior is similar for all classification strategies.

One very prominent result is a poor performance of the MIR classifier, which shows very low Cohen's kappa, Matthews correlation coefficient and precision results even in the case of the RBF kernel. This result indicates that even if the MIR is crucial to identify AGNs, the MIR data alone is not sufficient for this task. The better performance of the $nir$ classifier can be explained by the properties of the specific training samples. The reduced number of objects in the case of the $mir$ classifier could cause difficulties in the hyperplane construction process, not possible to overcome without additional information from the NIR passbands. The same tendency is visible in the ROC curve plot, where the $mir$ classifier curve is far from the top left corner.

The best performance results are obtained for the $6\_colors$ and $mi\_features$ classifiers. In the further analysis these classifiers will be considered as the most efficient.

\subsection{Extrapolation properties} \label{extrapol}

As it was mentioned in the previous section, the classifier trained on a particular data set will have difficulties in the extrapolation to the region outside the space occupied by the training data. Because of these difficulties the generalization sample should be limited to ranges of the training sample. However, trying to reflect original selection effects and due to difficulties in the proper representation of the high dimensional data it was decided to apply cuts only for flux measurements. Such cuts yield a more conservative limitation of the input feature space than cuts on the color values.

Limiting the space of the generalization sample only by introducing filter cuts has also additional implications. In this case high dimensional space occupied by the training sample is dipped in the generalization space which still corresponds to the bigger volume. Because of this issue the classifier will still have to extrapolate near the outskirts of the training data space in the generalization process.

To investigate the ability of the classifier to extrapolate outside the region of the training set, the training sample was divided into two parts: 70\% of the labeled data was assigned to a new training sample  and the remaining 30\%, containing objects with maximal (or minimal) values of the particular feature, were assigned to a new test sample. Such divisions were made separately for objects with the highest redshift and the faintest fluxes in the N2, S7 (or S9 in the case of $wide\_mir$ feature set) and L18 passbands. In particular differences in distributions of N2 flux and redshift values of training and generalization sets are shown in Fig.~\ref{fig:n2_properties}.

\begin{figure*}
\centering     
\subfigure[Evaluation metrics for all classifiers with the RBF kernel function. Values were obtained from the test sample of objects characterized by the highest redshift. Explanation the same as in Fig.~\ref{fig:rbfevalall}.]{\label{fig:redshift}\includegraphics[width=80mm]{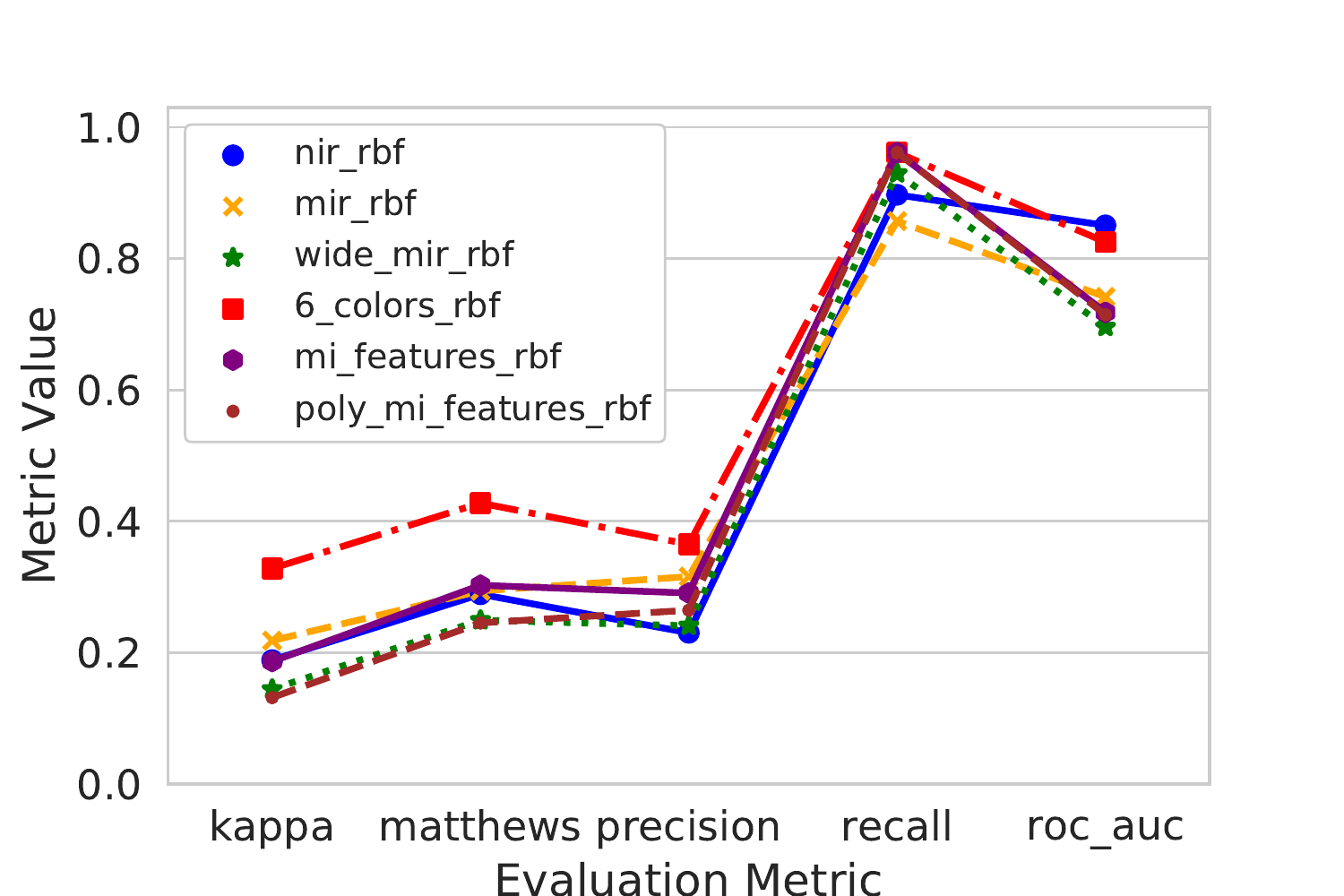}}
\quad
\subfigure[Evaluation metrics for all classifiers with RBF kernel function. Values were obtained from the test sample of the faintest objects in the N2 filter.]{\label{fig:N2}\includegraphics[width=80mm]{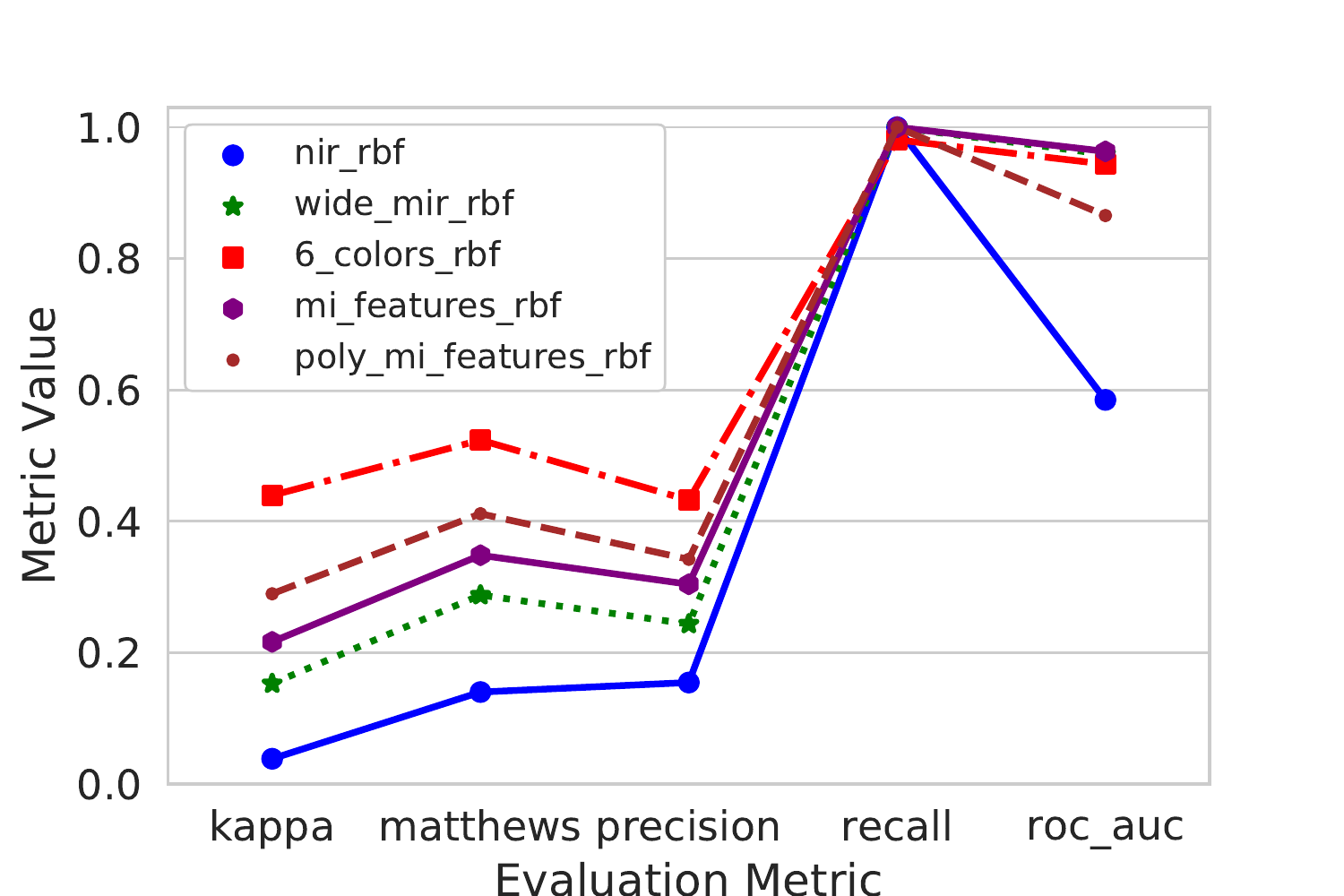}}

\subfigure[Evaluation metrics for all classifiers with RBF kernel function. Values were obtained from test sample of the faintest objects in S7 filter (S9 filter in the case of $mir$ feature set).]{\label{fig:S79}\includegraphics[width=80mm]{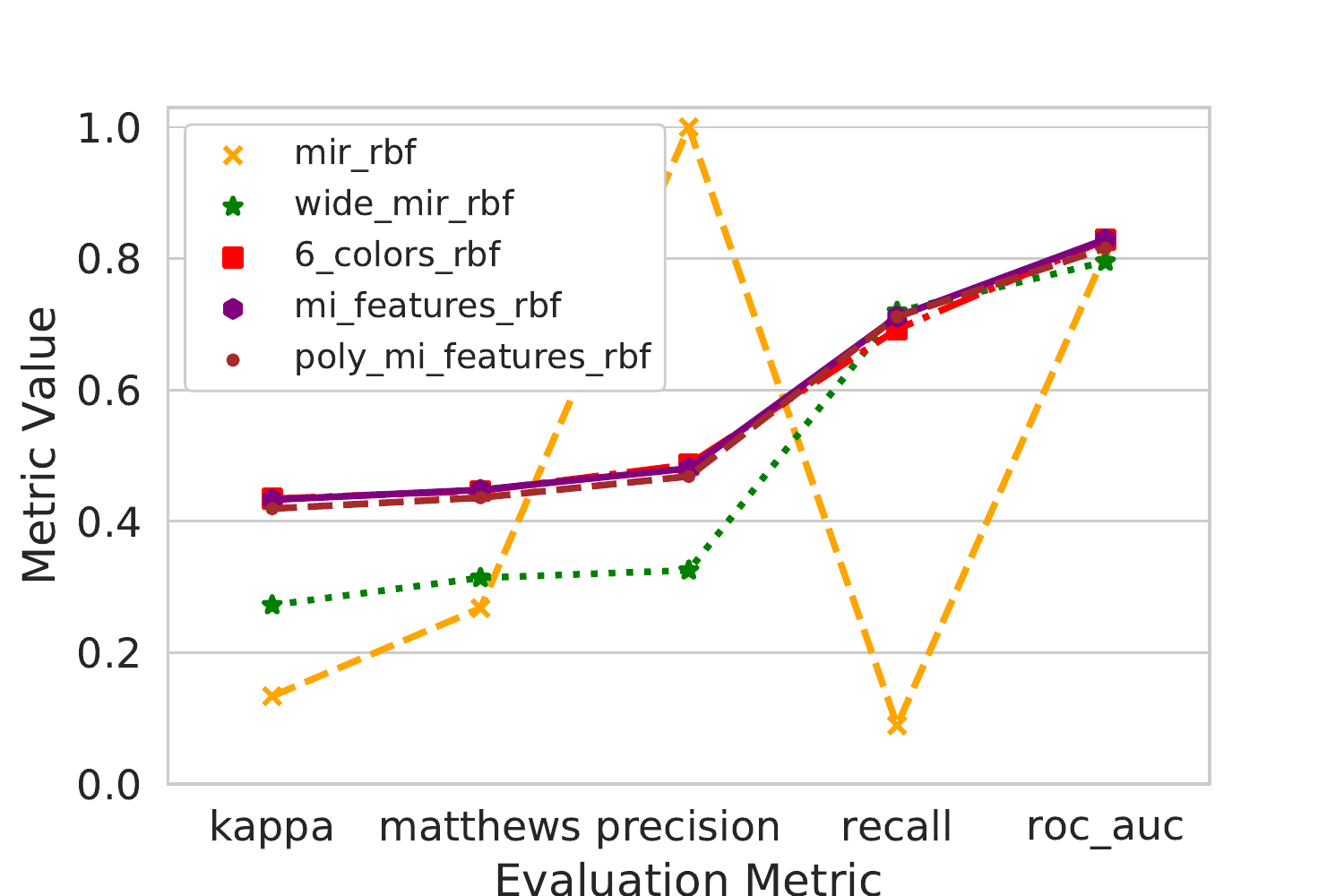}}
\quad
\subfigure[Evaluation metrics for all classifiers with RBF kernel function. Values were obtained from test sample of the faintest objects in L18 filter.]{\label{fig:L18}\includegraphics[width=80mm]{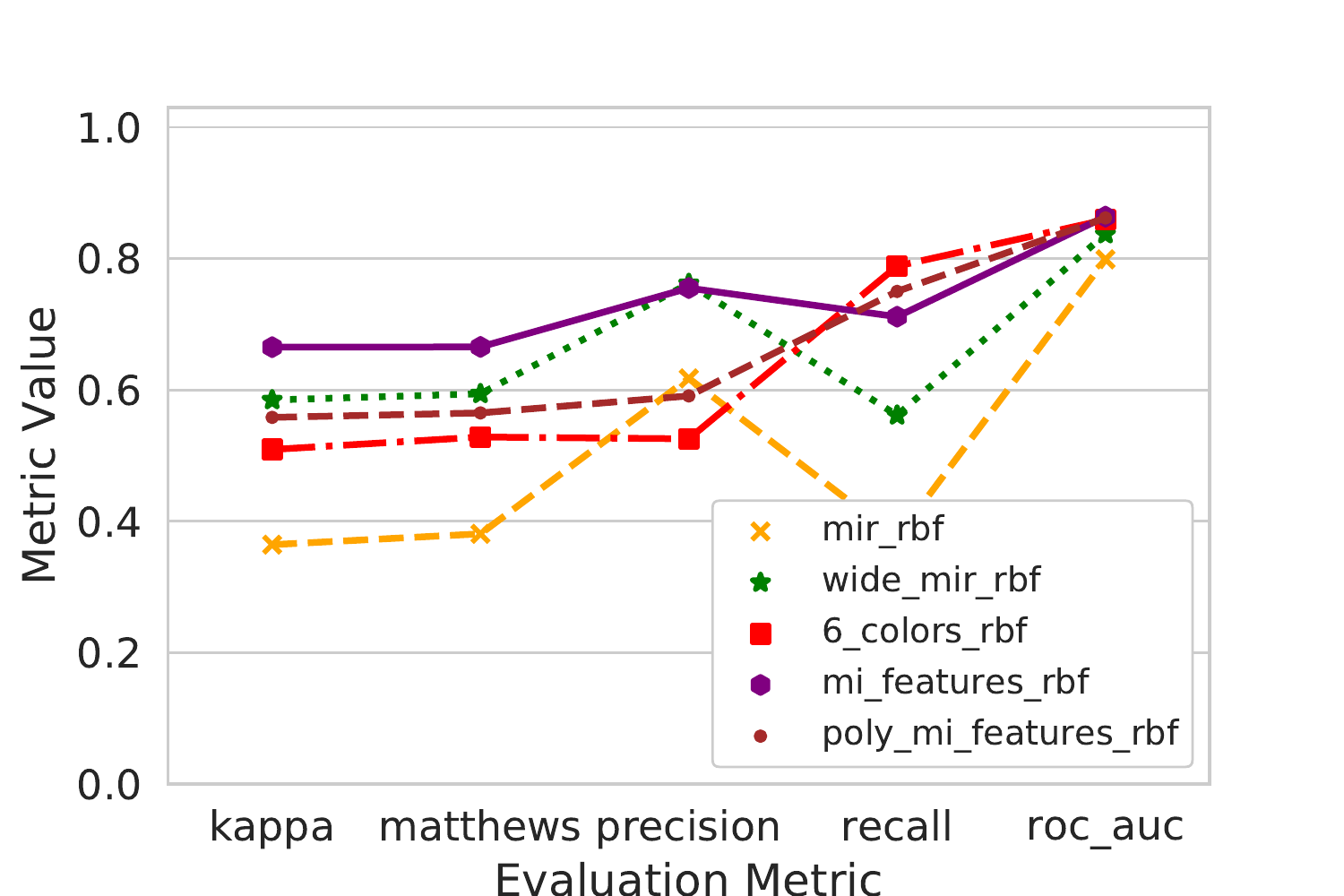}}
\caption{Extrapolation performance of the classifiers with RBF kernel trained on different feature sets. Table with exact evaluation metrics values can be found in Appendix 2.}
\label{fig:extrapolation}
\end{figure*}

After training the classifier on the new training set using 5-fold cross validation, its performance was tested on the extrapolation test set. Tables with exact results can be found in the Appendix~\ref{app_extrapol}. The $nir$ feature set classifier was excluded from the extrapolation tests with MIR limited passbands. The $mir$ feature set classifier was excluded from the extrapolation test with N2 limited passband. The visualization of the evaluation metric values for different extrapolation tests is presented in Figure~\ref{fig:extrapolation}.

Results shown in this section are only an approximation of the extrapolation efficiency of the classifier. The small amount of labeled data reduces the precision of the performance of the evaluation and constricts the investigation of the declining extrapolation performance cause. Because of these limitations one can only expect that the proper classifier will not show poor results in the extrapolation test. Performance on the redshift-limited test sample shows very high recall values with well preserved values of the ROC AUC score and drastically lower results in the case of precision, Cohen's kappa and Matthews correlation coefficient in comparison with the main evaluation discussed in Sect.~\ref{main_comparison}. High recall values and lower results for other evaluation metrics imply that classifier recognizes most of unlabeled objects as AGN candidates and, as a result, it will obtain a highly complete catalog with relatively low purity. In this case the $6\_colors$ classifier shows the best performance. A similar tendency is visible in the case of N2-limited sample. Results for the $nir$ classifier are drastically lower for this extrapolation experiment because of the lack of the MIR measurements, which could compensate limited NIR information. A similarly poor performance can be observed in the case of the $mir$ classifier for the experiments with S7 and L18-limited samples. These experiments show that in order to minimize the risk of the wrong classification in the case of extrapolation near the range of the training sample one has to use feature sets that contain both NIR and MIR measurements. Both classifiers selected in the main performance tests, $6\_colors$ and $mi\_features$ classifiers, show the best results in all tests except for the L18 limited experiment. Because of this fact, these two classifiers were selected as the best classifiers for the generalization task.

\subsection{Selection of the best classifier} \label{best_classifier}
Both $6\_colors$ and $mi\_features$ classifiers showed good, very similar results in the performance evaluation process. Finally, because of a better physical motivation, more convenient interpretation of the results and better performance in most of the extrapolation experiments, a $6\_colors$ classifier was chosen as the best model for the generalization.

\begin{figure*}
\centering     
\includegraphics[width=1.0\textwidth]{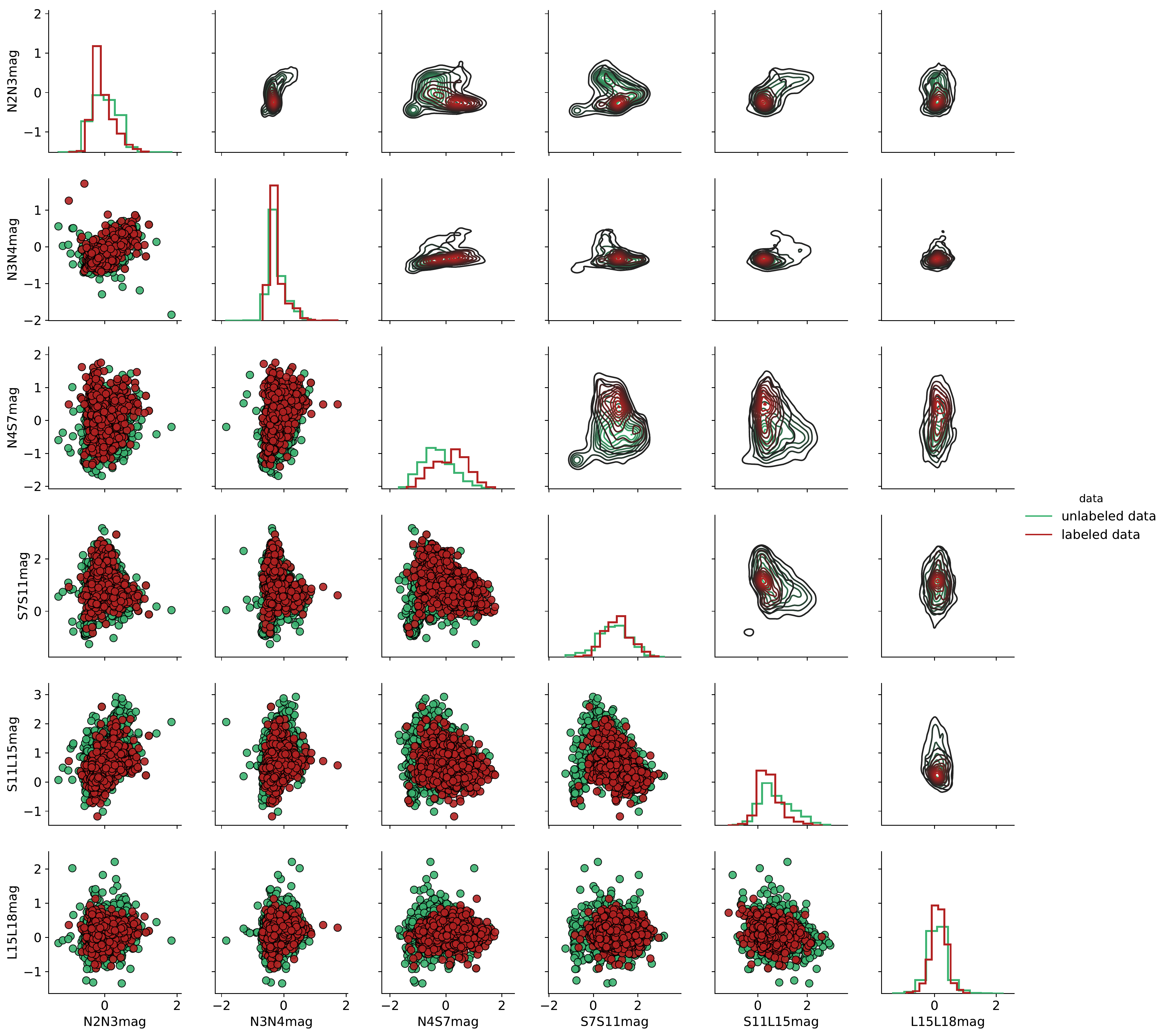}
\caption{Color properties of the unlabeled generalization data (green color) limited in particular filters to the range of the labeled training set (red color). Contour density plots were prepared using kernel density estimate. On the diagonal normalized histograms are shown.}
\label{fig:labeledunlabeledcolor}
\end{figure*}

Cuts applied to the generalization set were based on limitations of the $6\_colors$ training set used in Section~\ref{main_comparison}. These cuts are shown in Table~\ref{tab:range}. After they were applied, a generalization sample of 1716 objects was created. Color properties of the unlabeled generalization data in comparison with the training sample are shown in Figure~\ref{fig:labeledunlabeledcolor}. Distribution of fluxes in different NIR and MIR filters of the training and generalization $6\_color$ feature set samples, corresponding flux uncertainties and fuzzy memberships for labeled data can be found in Appendix~\ref{app_6col}.

\begin{table*}[t]
\tbl{The range of the $6\_color$ feature set training sample, applied for the final catalog for generalization. Columns show minimal and maximal values of colors used for the construction of the input feature space.}{%
\begin{tabular}{lcccccc}
\hline
Limit & N2N3 [mag] & N3N4 [mag] & N4S7 [mag] & S7S11 [mag] & S11L15 [mag] & L15L18mag\\
min & -0.99 & -1.85 & -1.68 & -1.26 & -1.02 & -1.34\\
max &  1.22 &  0.86 &  1.60 &  3.17 &  2.93 &  2.21\\
\hline
\end{tabular}}\label{tab:range}
\end{table*}

The distribution of most of colors have similar ranges in the labeled and unlabeled sets. However, shapes of the MIR color distributions are different. It can be caused by the differences in properties of samples from which they were drawn. The training set is mostly made of objects with optical counterparts from a shallower survey (AKARI NEP Wide), while the generalization set preserved objects with no optical counterparts. Consequently, as a sample from deeper survey (AKARI NEP Deep), it contains higher redshift objects with different properties and with different lines located in the MIR range.

\section{Results}

\subsection{Software}

Parameters of the best RBF classifier (presented in Table~\ref{tab:evaluation}) trained on the $6\_colors$ feature set making use of the whole training data as it was described in~\ref{main_comparison} were chosen to be $C=500$ and $\gamma=0.0015$ in the grid search process. Time needed for training of each classifier was equal to approximately 20-60 seconds on the machine with 16 GB RAM and Intel i7-4790K. The generalization takes approximately 10 seconds. The code used in the present work is available from the github repository: github.com/ArtemPoliszczuk.

\subsection{AGN catalog}

\begin{figure*}
\centering     
\subfigure[Distance from the separating hyperplane of the AGN candidates (blue) and Other class candidates (orange).]{\label{fig:distance}\includegraphics[width=80mm]{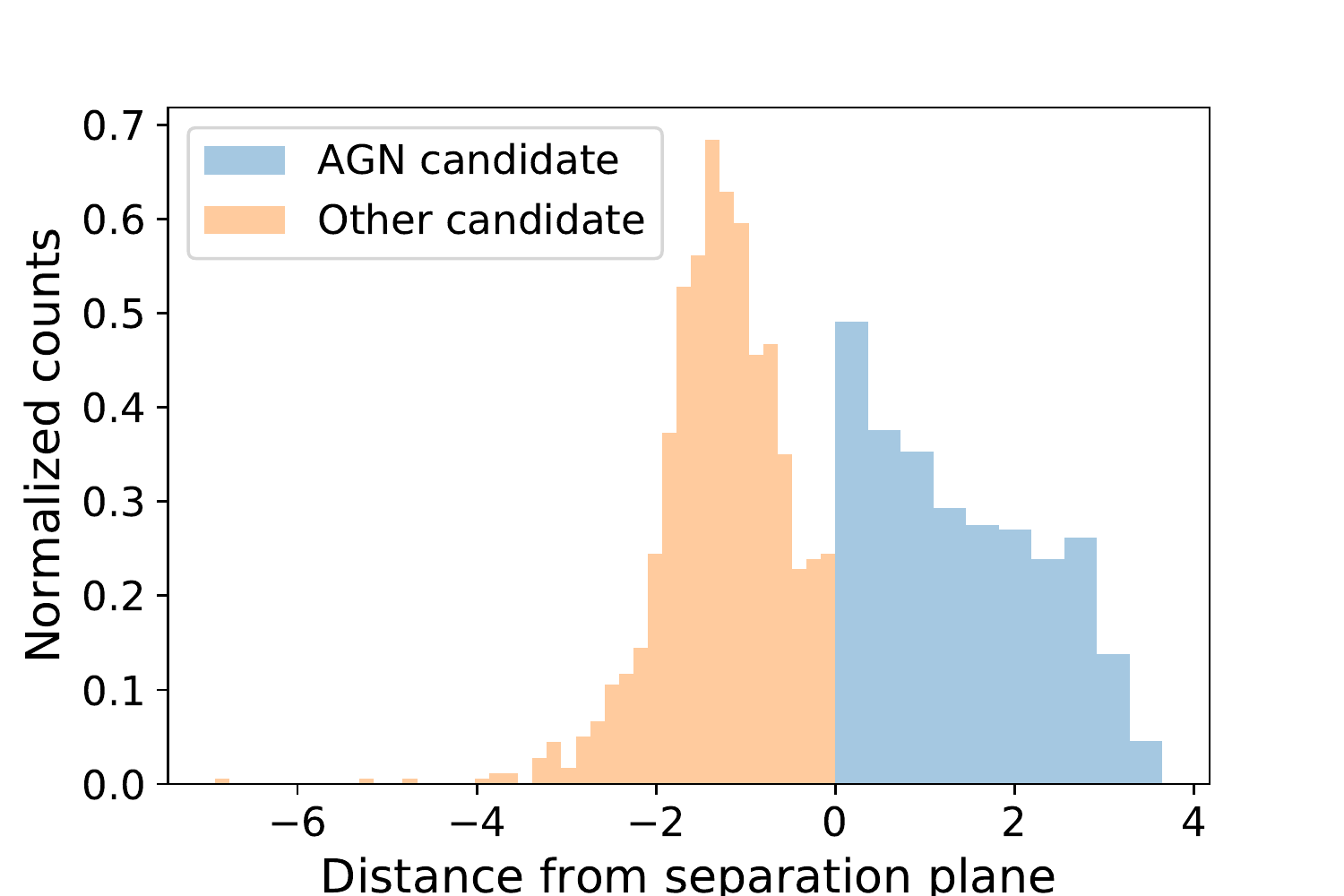}}
\qquad
\subfigure[Platt's probability of being an AGN for the AGN candidates (blue) and Other class candidates (orange).]{\label{fig:prob}\includegraphics[width=80mm]{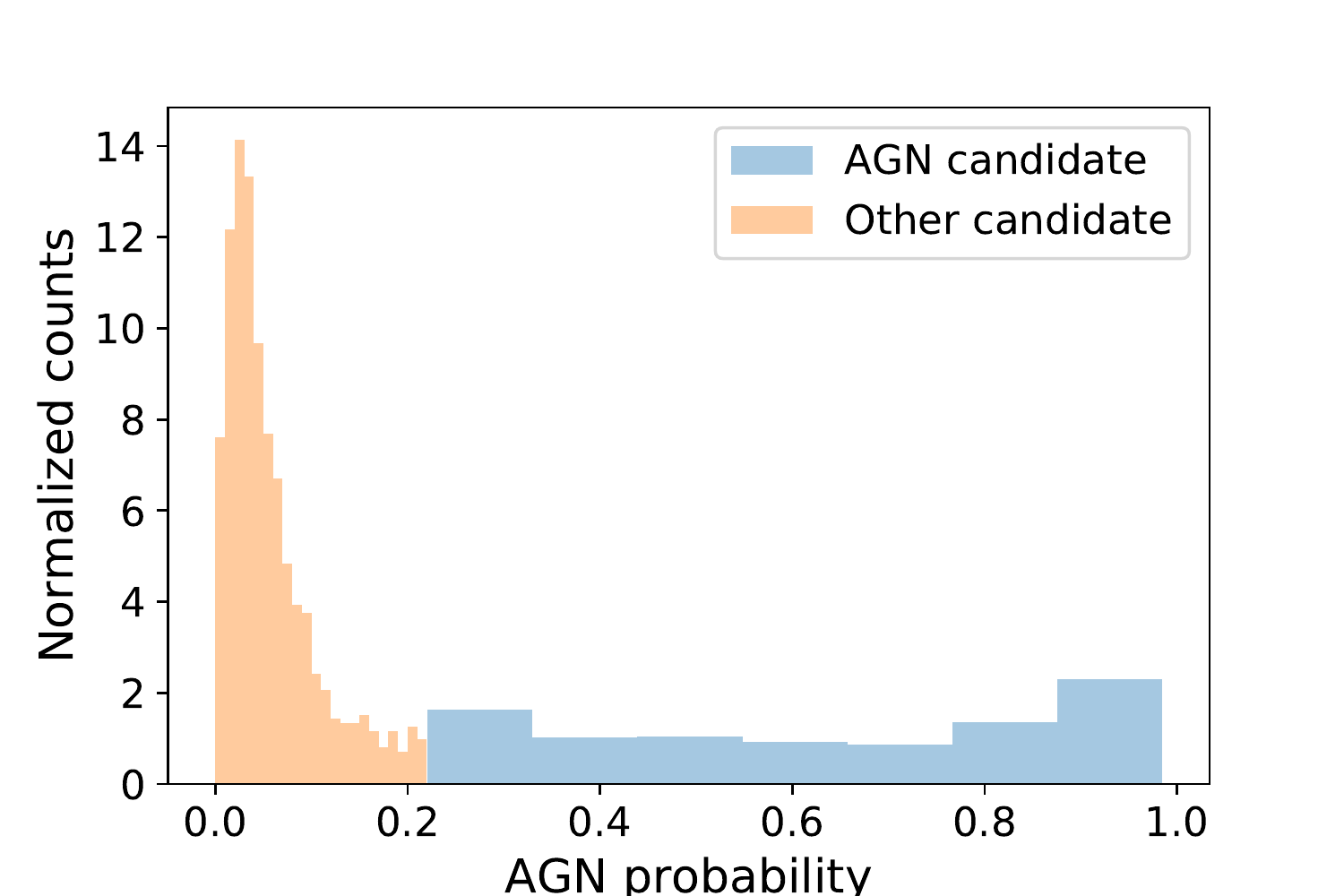}}
\caption{Distance from the separating hyperplane and corresponding Platt's probability for the candidates of AGN and Other classes}\label{fig:distprob}
\end{figure*}

Using the $6\_colors$ classifier a set of 598 AGN candidates and 1118 candidates of the "Other" class was obtained. To further analyze the quality of AGN candidates catalog one can plot the distribution of the distance of objects from the decision surface and corresponding Platt's probability of an object being properly classified. Both of these distributions are shown in Figure~\ref{fig:distprob}.

\begin{figure*}
\centering     
\subfigure[Platt's probability of being an AGN for the AGN candidates.]{\label{fig:agnprob_agnonly}\includegraphics[width=80mm]{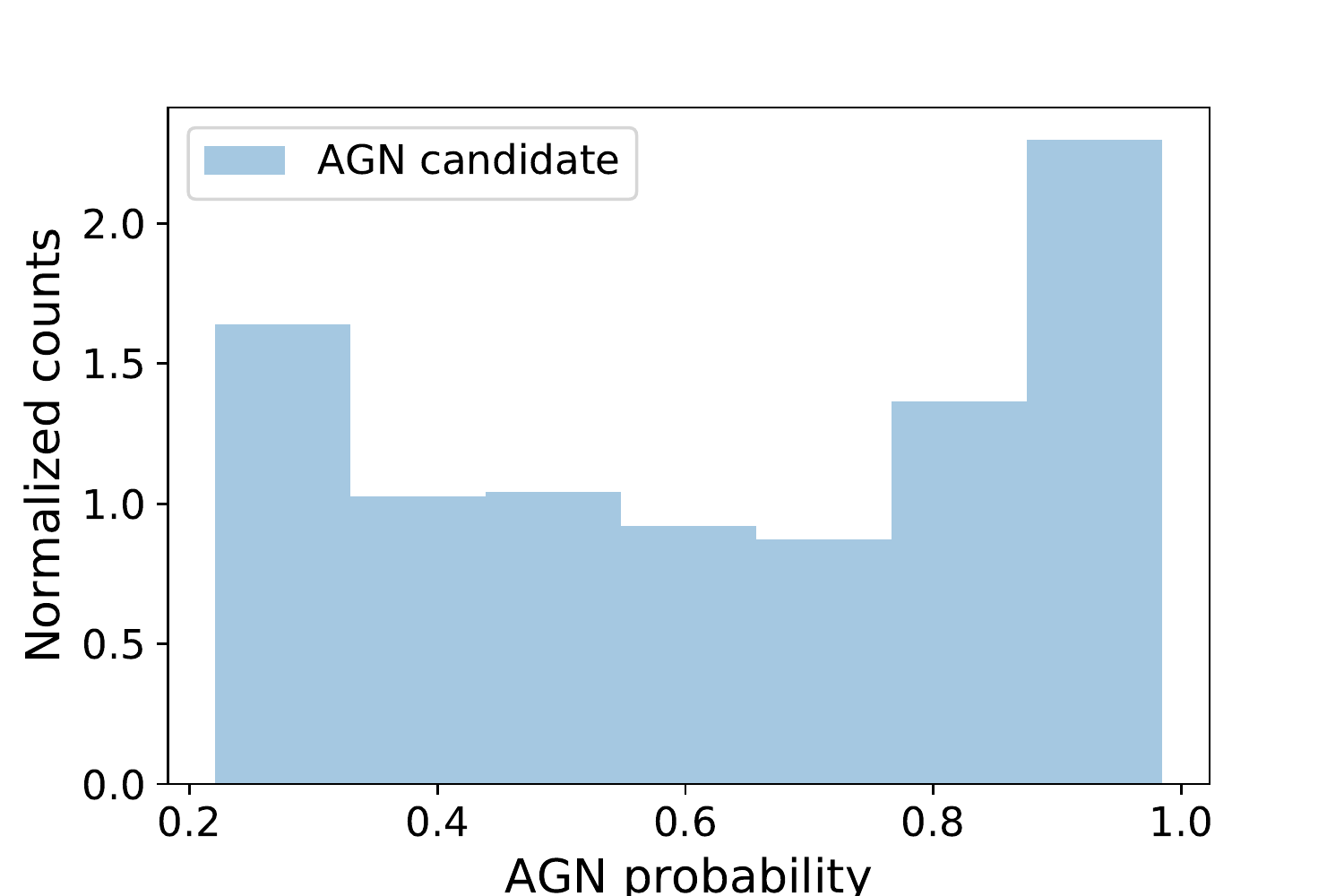}}
\qquad
\subfigure[Relation between the probability cut and the number of objects remaining in the catalog of AGN candidates.]{\label{fig:relation_prob_num}\includegraphics[width=80mm]{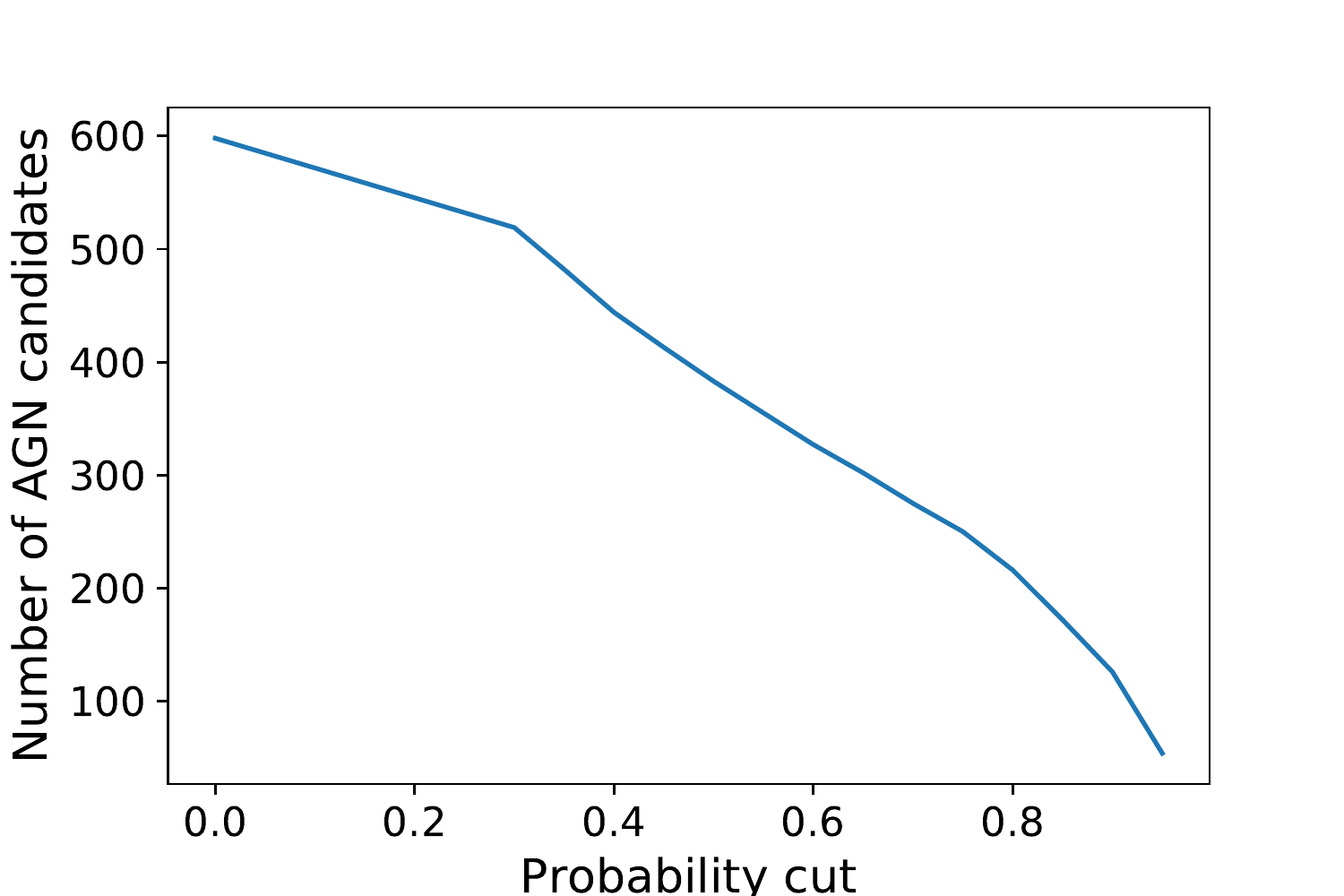}}
\caption{Platt's probability of being an AGN for the AGN candidates and the relation between the number of objects remaining in the catalog of AGN candidates and the probability cut threshold.}\label{fig:agnprob}
\end{figure*}

Probability values lower than 0.5 in the case of the AGN candidates occur because of imbalanced sizes of training sets. The SVM has a tendency to shift the separating hyperplane towards the smaller class~\citep{batuwita12}. As a consequence, more objects from the smaller class lie near the boundary. Due to the direct relation between the distance from the decision surface and the Platt's probability, this shift results in probability values lower than 0.5. In such a situation one cannot treat values of the Platt's probability as directly applicable measures. However, the probability distribution still can be used to select objects with the highest confidence of a particular class membership. Figure~\ref{fig:agnprob} shows the Platt's probability distribution only for AGN candidates and the relation between the probability cut and the number of objects remaining in the catalog of AGN candidates.

\begin{figure*}
\centering     
\includegraphics[width=1.0\textwidth]{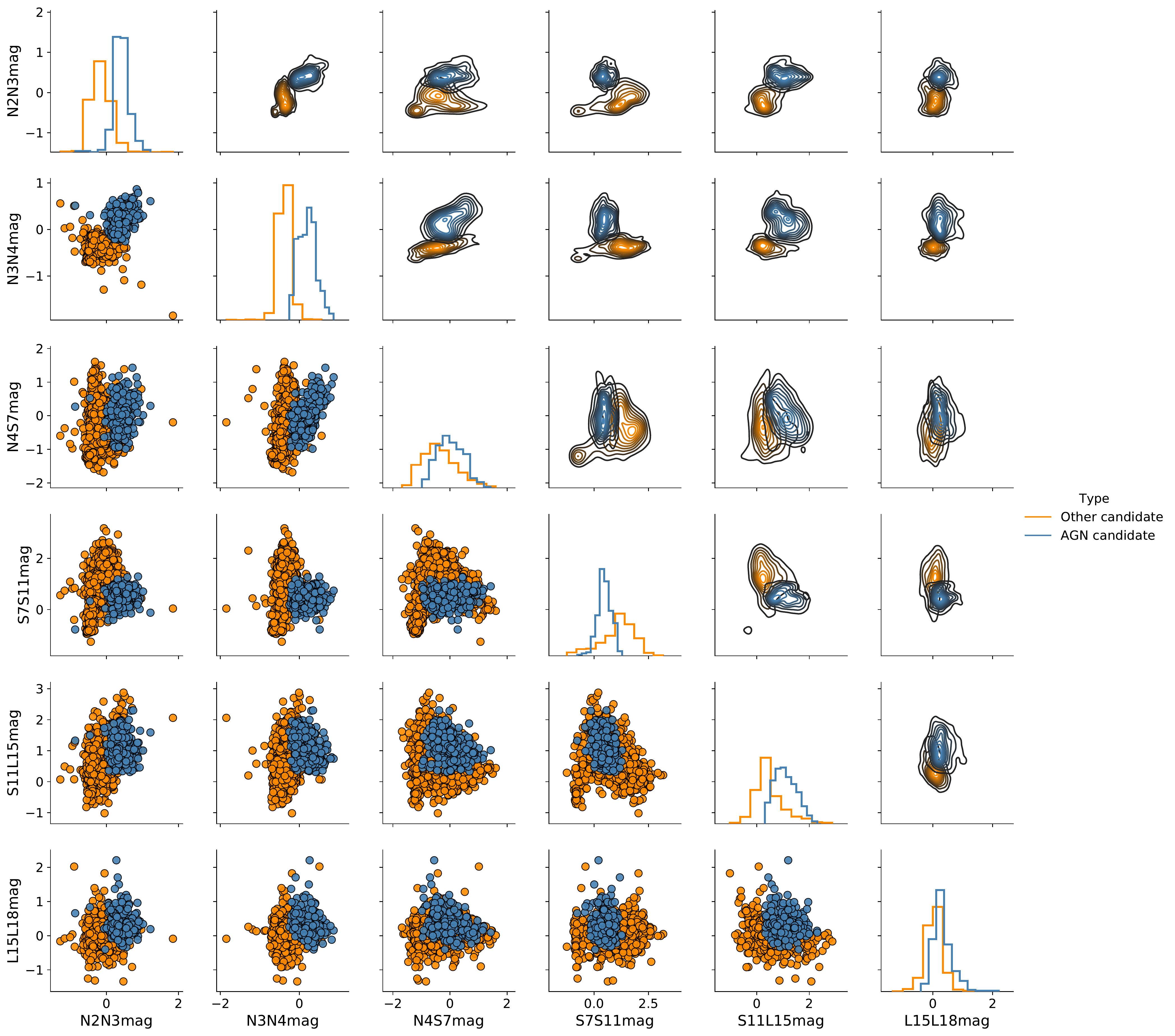}
\caption{Color properties of the final AGN candidates catalog (blue color) and objects that were classified as 'Other' class (orange color).Contour density plots were prepared using kernel density estimate. On the diagonal normalized histograms are shown.}
\label{fig:colors_final}
\end{figure*}

\begin{figure*}
\centering     
\subfigure[Histogram of $N2-N4$ color for AGN candidates with different probability threshold.]{\label{fig:n2n4agnhistprob}\includegraphics[width=80mm]{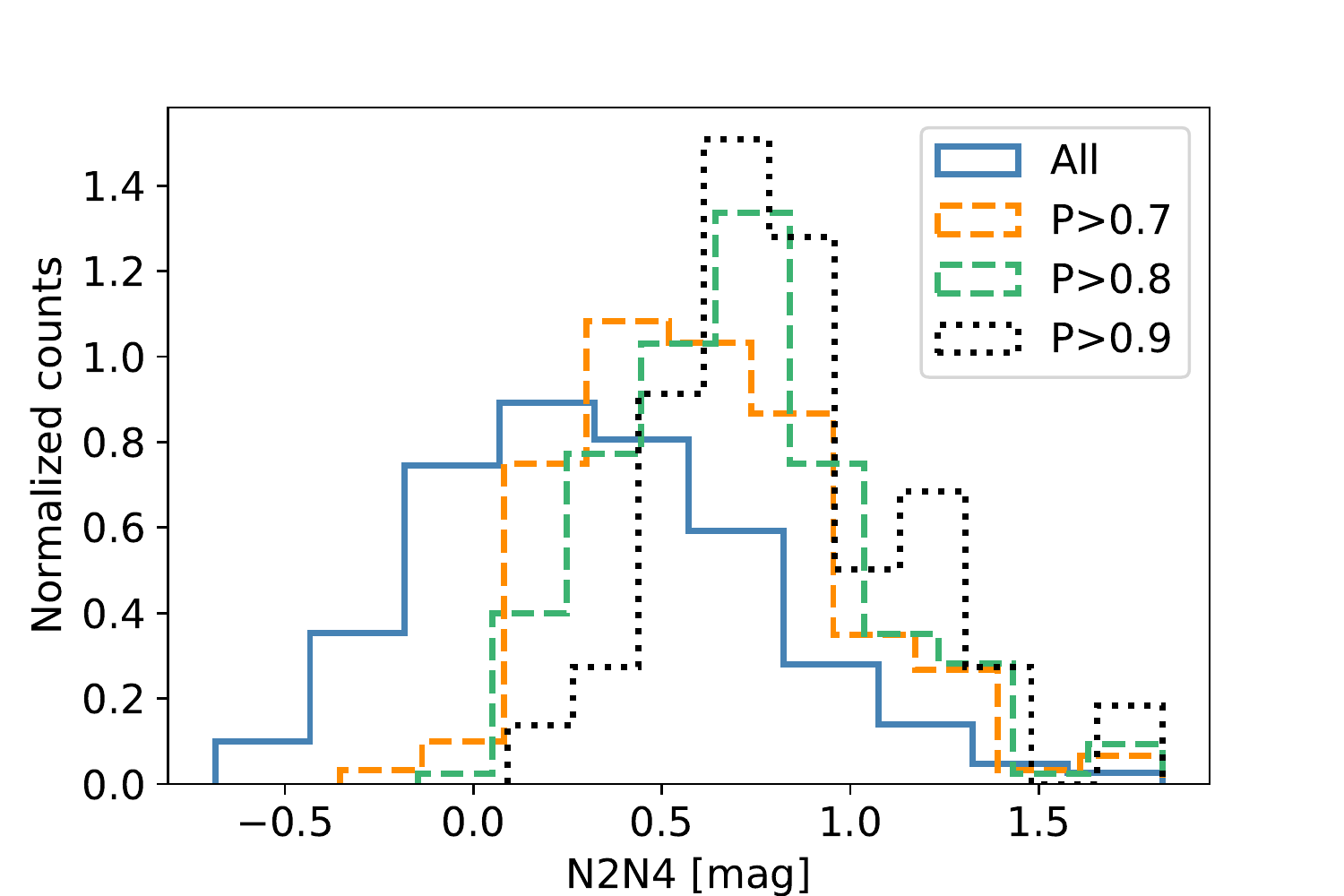}}
\qquad
\subfigure[Histogram of $S7-S11$ color for AGN candidates with different probability threshold.]{\label{fig:s7s11agnhistprob}\includegraphics[width=80mm]{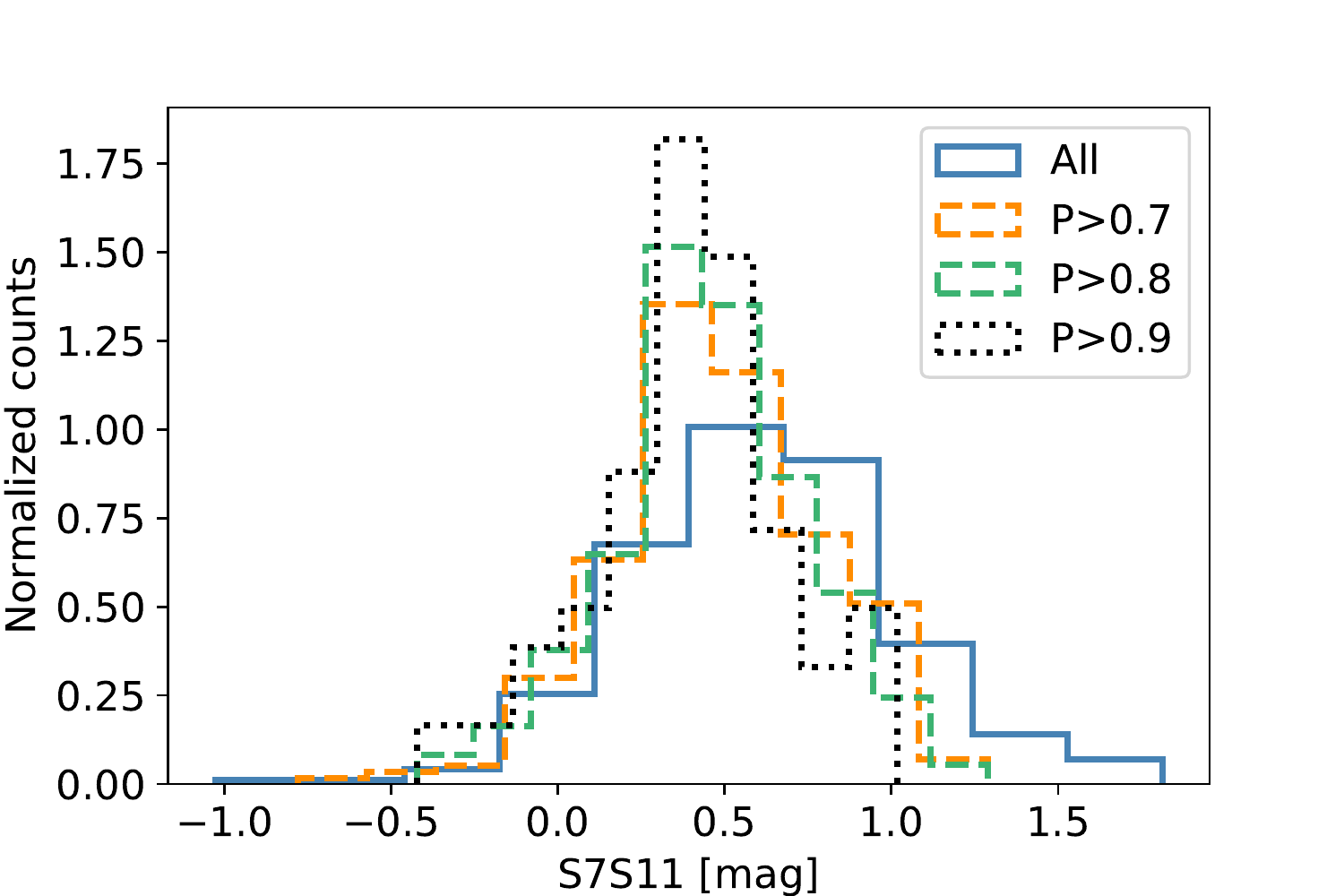}}
\caption{Color histograms for AGN candidates with different probability threshold. All AGN candidates without any probability cut are referred to as "All" objects. Final catalog of AGN candidates was made of objects with probability threshold $P>0.7$.}
\label{fig:agnhis_different_prob}
\end{figure*}

The probability threshold was set on the value $P=0.7$ in the minimum before the peak of the objects with the highest confidence value. As a result, a catalog of 275 AGN candidates was obtained. None of these candidates was present in the training sample. Color properties of the final AGN candidates catalog are shown in Figure~\ref{fig:colors_final}.
Additional histograms of $N2-N4$ and $S7-S11$ colors for AGN candidates with different probability thresholds are shown in Figure~\ref{fig:agnhis_different_prob}. It is seen that distributions shift with the change of the probability threshold. In the case of $N2-N4$ color the distribution is shifted towards redder values, while in the case of the $S7-S11$ color distribution moves towards bluer values.  These two particular colors were selected as they allow for comparison with other AGN selection method which is presented in the next Section.

\subsection{Comparison with color selection method}

\citet{lee07} studied properties of sources brighter than 18.5 mag at 11$\mu$m in the Early AKARI NEP-Deep data and proposed an AGN selection method based on the $N2$-$N4$ vs $S7$-$S11$ color-color diagram, where AGNs occupy an area of $N2$-$N4>$0 and $S7$-$S11>$0.

\begin{figure}
\centering     
\includegraphics[width=0.5\textwidth]{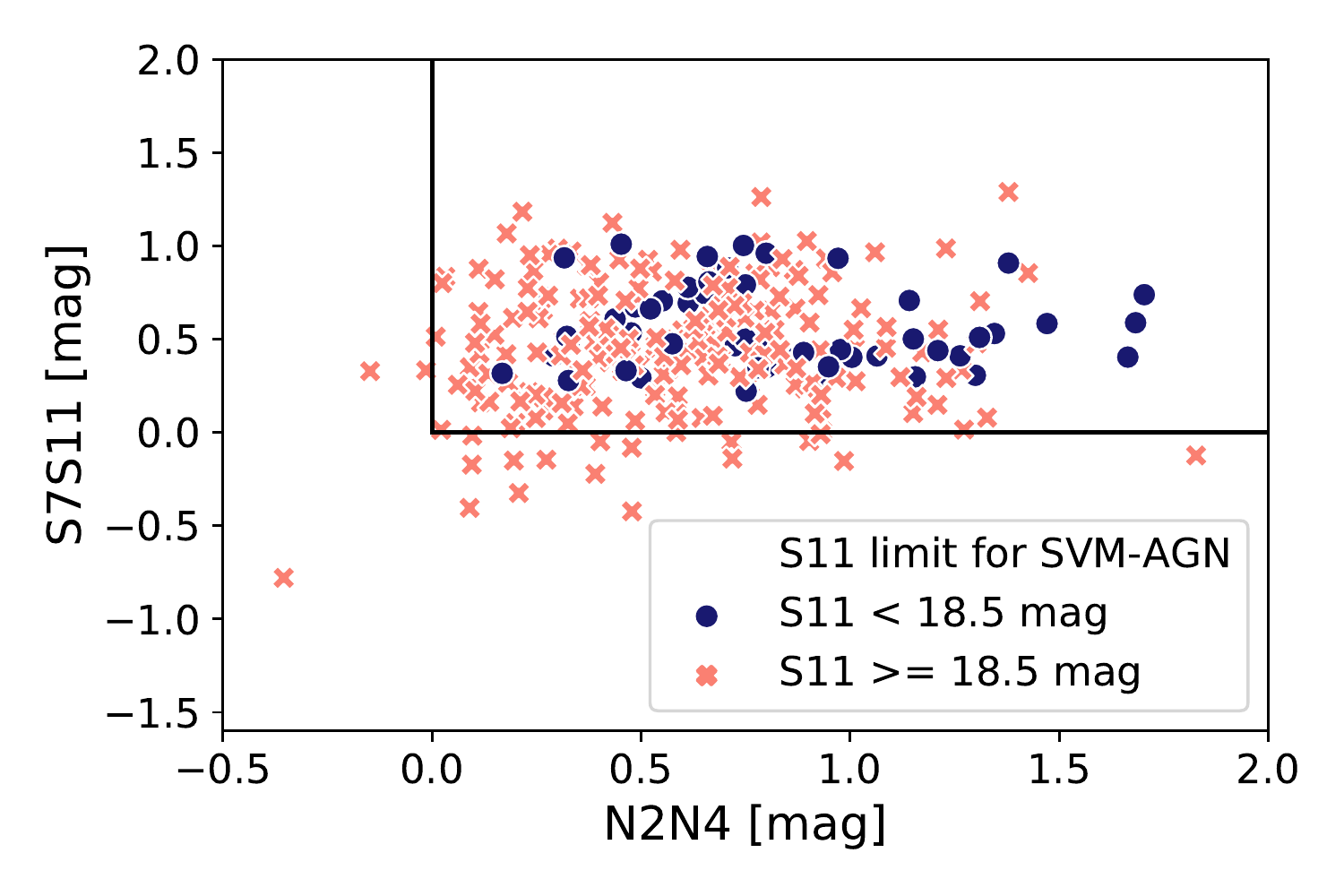}
\caption{$N2$-$N4$ vs $S7$-$S11$ color-color diagram of SVM-selected AGN candidates (blue circles for objects brighter than 18.5 mag at 11$\mu$m and pink crosses for fainter objects) together with the cuts proposed by~\citet{lee07} (upper right corner restricted by black lines).}
\label{fig:n2n4_s7s11}
\end{figure}

Among SVM-selected AGN candidates with probability cut applied, 256 (93\% of the sample) fulfill this criterion, while 36 objects (12\% of the generalization sample) were not classified as AGNs. However, if the original $S11 < 18.5 mag$ cut is applied to all the samples, the numbers shrink to 55 (100\% of the sample) fulfilling the~\citet{lee07} criterion, and 17 objects (3\% of the generalization sample) not classified as AGNs with high probability by SVM in spite of fulfilling the~\citet{lee07} criterion. The $N2$-$N4$ vs $S7$-$S11$ color-color diagram of SVM-selected AGN candidates, together with the cuts proposed by~\citet{lee07} is presented in Fig.~\ref{fig:n2n4_s7s11}. One can conclude that both methods are largely consistent but do not yield exactly the same selection; multi-color feature space very likely allows for a more robust classification. To demonstrate that, the properties of the objects with different classification by both methods will be analyzed in the future work.

\section{Summary}
In the present work the fuzzy SVM algorithm with the fuzzy memberships based on the measurement uncertainties was applied to the AKARI NEP-Deep data for the AGN selection task. Six different feature sets were tested. Some of them were physically motivated, some were constructed using filter feature selection method. Classifiers based on different kernels and feature sets were tested by using a set of evaluation metrics. Additional extrapolation experiments were also performed. The feature set made of six colors of neighboring passbands turned out to be the most efficient and informative. The classifier based on the RBF kernel function and 6 colors feature set was then used for the creation of the AGN candidates catalog. The final AGN catalog was made by applying additional probability cuts in order to obtain objects with the most reliable classification outcome. The majority of sources (77\%) selected as AGN candidates by SVM fulfills the simpler color-color selection criterion proposed by~\citet{lee07} but the selection is not exactly the same. The properties of the deviating sources will be studied in the future work.

AKARI NEP-Deep data turned out to be a challenging ground for the automated classification tasks. The shallower training sample in comparison with the data used for generalization increases the risk of the bad representativeness of the labeled data - this is the problem of most of supervised classifications performed on the photometric survey data with labels taken from spectroscopic surveys. Additional small volume of such labeled sample makes the differences between the training and generalization samples harder to control. As a consequence, more advanced tools, such as extrapolation experiments and Platt's probability thresholds were applied in order to increase the reliability of the output catalog.

Because of many different selection effects and probability cut, the completeness of the final catalog was strongly reduced. The main purpose of this work was to obtain a reliable catalog of AGNs characterized by the high purity. This goal was accomplished, however, it is hard to precisely evaluate purity of the final catalog. A simple look on the precision value cannot be treated as a real purity measure due to the imperfect representativeness of the training sample and additional purity raise after the probability cut.

The key role in the creation of the final catalog was played by the physical information incorporated into the classification process both by applying a fuzzy memberships based on measurement uncertainties and using well physically motivated feature set. Such a construction of fuzzy memberships allowed to increase the importance of properly measured objects in the process of decision boundary creation and to reduce the influence of the artifacts specific for the current data set.

The new catalog of AGN candidates contains a potentially unique set of faint infrared AGNs and will be investigated in the further research.

 \begin{ack}
 Authors are very thankful to the referee for corrections, which allowed to significantly improve the manuscript.\newline
 Authors are grateful for the support from the Polish Ministry of Science and Higher Education through a grant DIR/WK/2018/12.\newline
 A. Poliszczuk, A. Pollo and M. Bilicki were  supported  by  the Polish National  Science  Centre  grant  UMO-2012/07/D/ST9/02785. \newline
A. Poliszczuk was supported by the Polish Ministry of Science and Higher Education grant MNiSW 212727/E-78/M/2018. \newline
A. Pollo was supported by Polish National Science Centre, grant No. 2017/26/A/ST9/00756. \newline
M.B. was supported by the Netherlands Organization for Scientific Research, NWO, through grant number 614.001.451. \newline
 A. Solarz was  supported  by  the Polish National  Science  Centre  grant  UMO-2015/16/S/ST9/00438. \newline
 Tomotsugu Goto acknowledges the support by the Ministry of Science and Technology of Taiwan through grant 105-2112-M-007-003-MY3. \newline
 T. Miyaji is supported by UNAM-DGAPA-PAPIIT IN111319 and CONACyT 252531.
 
 \end{ack}

\onecolumn

\appendix

\section{Results of the extrapolation experiments}
\label{app_extrapol}
Results of the extrapolation experiments described in Sect.~\ref{extrapol}. Table~7 shows values of different evaluation metrics for a sample of 30\% of labeled data with extreme values of particular flux or redshift, which was excluded from the training process. The best results were obtained by $6\_colors$ and $mi\_features$ RBF classifiers.


\begin{table*}[h]
\caption{Evaluation of different classifiers in the extrapolation experiment. In columns 2-6 values of different evaluation metrics (ROC AUC, which was maximized in the grid search process, Cohen's kappa, Matthews correlation coefficient, precision and recall) are shown. Evaluation results were obtained using a sample of 30\% of labeled data with extreme values of particular flux or redshift, which was excluded from the training process.}
\centering
\subfigure[Results for sample of objects with the highest corresponding redshift.]{
   \begin{tabular}{lccccc}
  \hline
	kernel    & roc auc         & kappa           & matthews        & precision       & recall \\
  \hline
nir\_rbf                & 0.85 & 0.19 & 0.29 & 0.23 & 0.90\\
mir\_rbf                & 0.74 & 0.22 & 0.29 & 0.32 & 0.85\\
wide\_mir\_rbf          & 0.70 & 0.14 & 0.25 & 0.24 & 0.93\\
6\_colors\_rbf          & 0.82 & 0.33 & 0.43 & 0.36 & 0.96\\
mi\_features\_rbf       & 0.72 & 0.19 & 0.30 & 0.29 & 0.96\\
poly\_mi\_features\_rbf & 0.71 & 0.13 & 0.24 & 0.26 & 0.96\\
\hline
\end{tabular}}\label{tab:evaluation_z}
\subfigure[Results for sample of the faintest objects in the N2 passband.]{
 \begin{tabular}{lccccc}
  \hline
	kernel    & roc auc         & kappa           & matthews        & precision       & recall \\
  \hline
nir\_rbf                & 0.59 & 0.04 & 0.14 & 0.15 & 1.0\\
wide\_mir\_rbf          & 0.96 & 0.15 & 0.29 & 0.24 & 1.0\\
6\_colors\_rbf          & 0.94 & 0.44 & 0.52 & 0.43 & 0.98\\
mi\_features\_rbf       & 0.96 & 0.22 & 0.35 & 0.30 & 1.0\\
poly\_mi\_features\_rbf & 0.86 & 0.29 & 0.41 & 0.34 & 1.0\\
\hline
\end{tabular}}\label{tab:evaluation_N2}
\quad
\subfigure[Results for sample of the faintest objects in the S7 (or S9 in the case of \textit{wide\_mir}) passband.]{
   \begin{tabular}{lccccc}
  \hline
	kernel    & roc auc         & kappa           & matthews        & precision       & recall \\
  \hline
mir\_rbf                & 0.81 & 0.13 & 0.27 & 1.0 & 0.09\\
wide\_mir\_rbf          & 0.79 & 0.27 & 0.31 & 0.32 & 0.72 \\
6\_colors\_rbf          & 0.83 & 0.43 & 0.45 & 0.49 & 0.69\\
mi\_features\_rbf       & 0.83 & 0.43 & 0.45 & 0.48 & 0.71\\
poly\_mi\_features\_rbf & 0.82 & 0.42 & 0.44 & 0.47 & 0.71\\
\hline
\end{tabular}}\label{tab:evaluation_S79}
\quad
\subfigure[Results for sample of the faintest objects in the L18 passband.]{
  \begin{tabular}{lccccc}
  \hline
	kernel    & roc auc         & kappa           & matthews        & precision       & recall \\
  \hline
mir\_rbf                & 0.80 & 0.36 & 0.38 & 0.62 & 0.37\\
wide\_mir\_rbf          & 0.84 & 0.58 & 0.59 & 0.76 & 0.56\\
6\_colors\_rbf          & 0.86 & 0.51 & 0.53 & 0.52 & 0.79\\
mi\_features\_rbf       & 0.86 & 0.66 & 0.66 & 0.75 & 0.71\\
poly\_mi\_features\_rbf & 0.86 & 0.56 & 0.56 & 0.59 & 0.75\\
\hline
\end{tabular}}\label{tab:evaluation_L18}
\end{table*}\label{tab:extrapolation_experiment}


\clearpage

\section{Properties of the data used for the 6 colors feature set} \label{app_6col}

\begin{figure}[h!]
\centering     
\subfigure[N2 and N3 flux distribution of the objects from the generalization unlabeled sample and training sample. ]{\label{fig:evalall}\includegraphics[width=80mm]{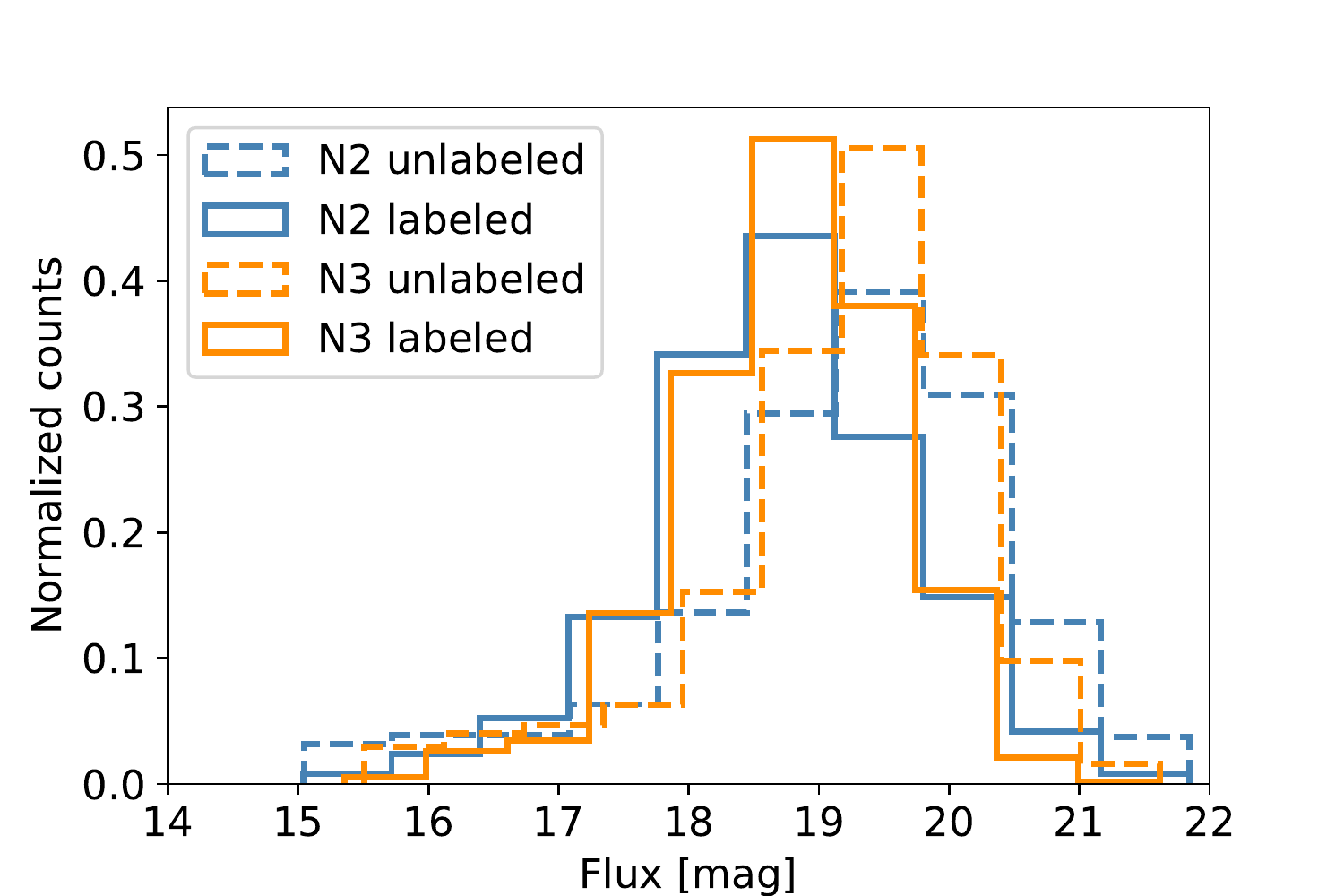}}
\quad
\subfigure[N2 and N3 flux measurement uncertainty distribution for objects from training and generalization samples.]{\label{fig:a}\includegraphics[width=80mm]{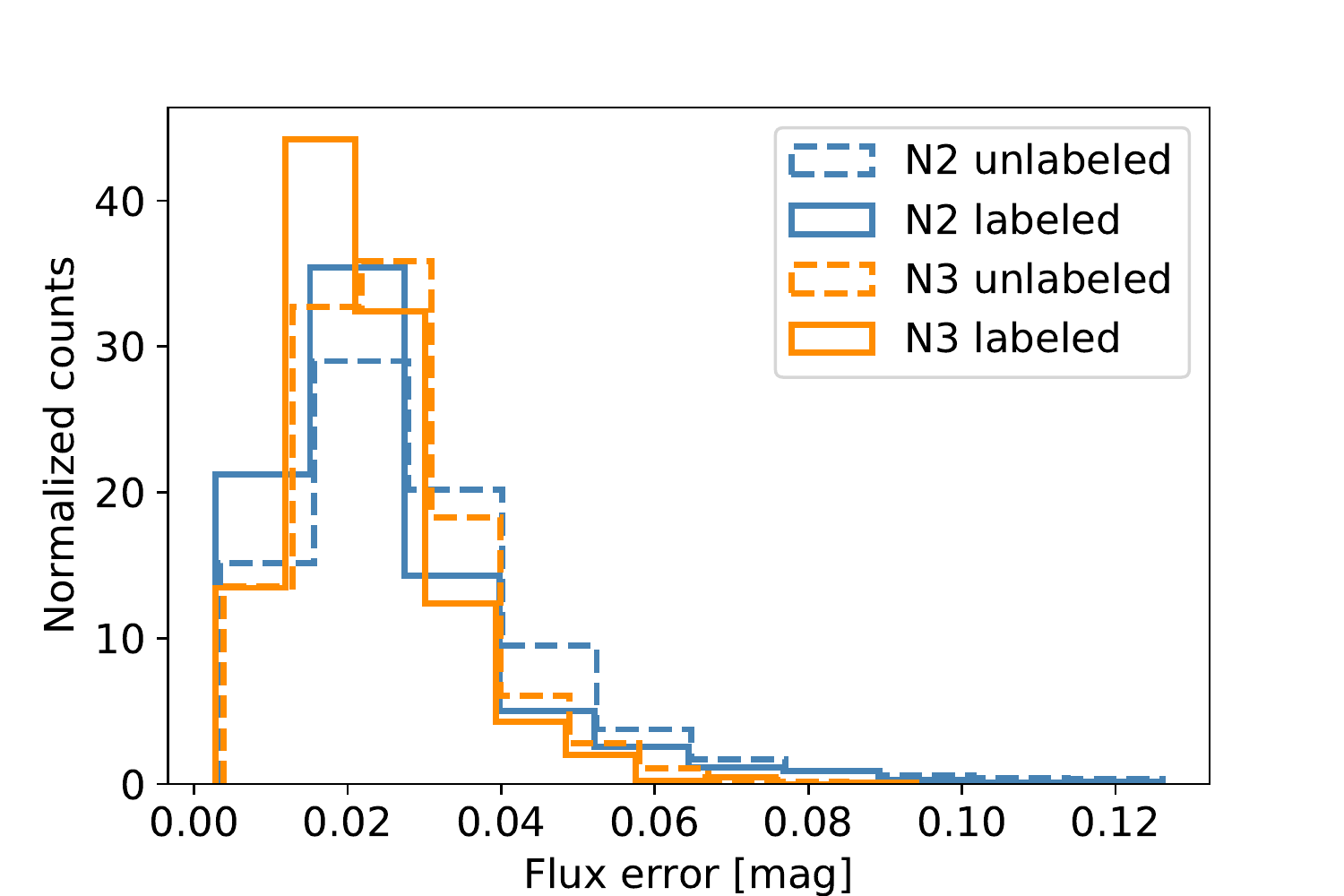}}
\caption{Distribution of the flux measurements and its uncertainties in N2 and N3 passbands for the $6\_color$ feature set data used for training in Sect.~\ref{main_comparison} and for generalization in Sect.~\ref{best_classifier}.}
\label{fig:fluxN2N3_6col}
\end{figure}

\begin{figure}[h!]
\centering     
\subfigure[N4 flux distribution of the objects from the generalization and training samples.]{\label{fig:a}\includegraphics[width=80mm]{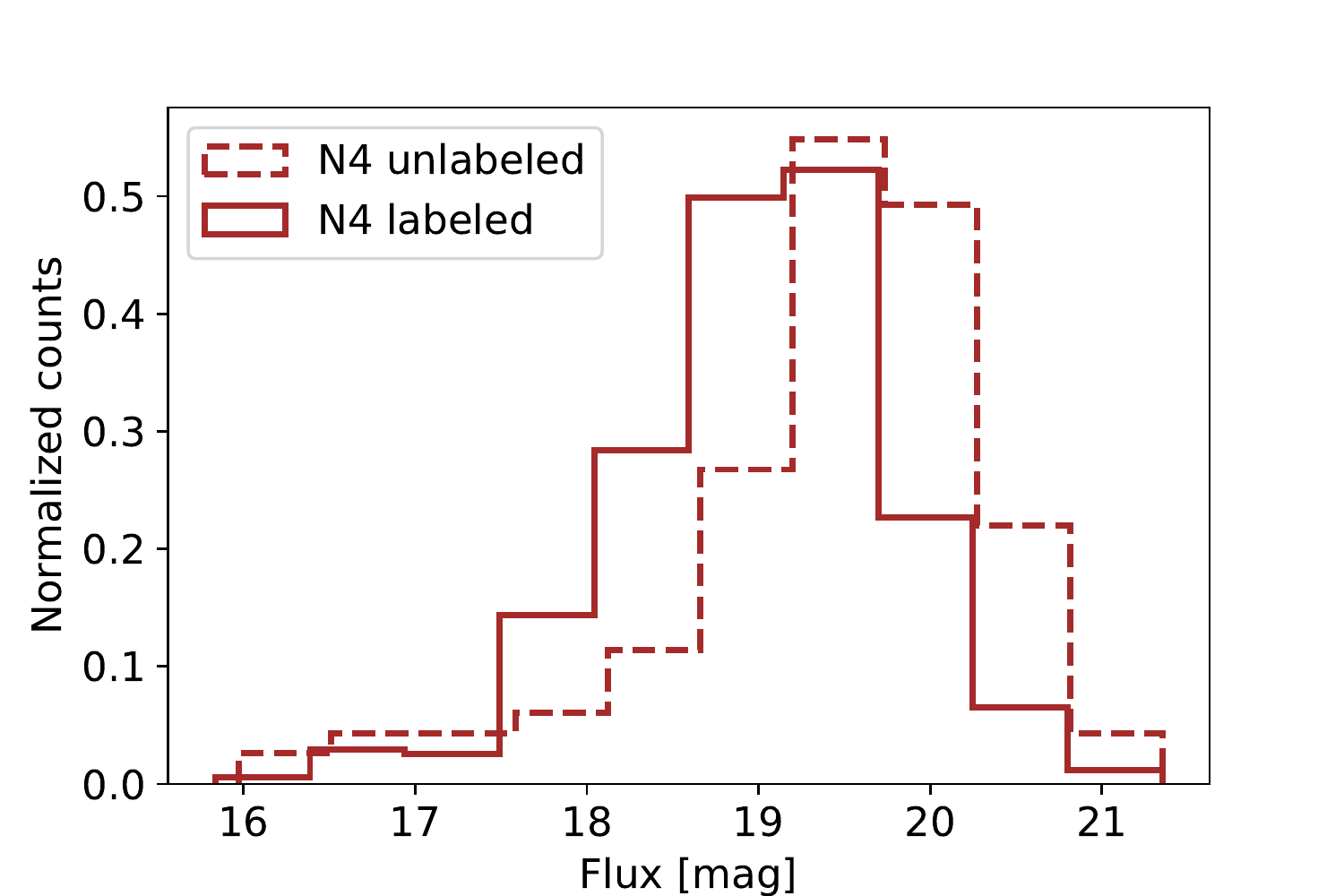}}
\quad
\subfigure[N4 flux measurement uncertainty distribution for objects from training and generalization samples.]{\label{fig:a}\includegraphics[width=80mm]{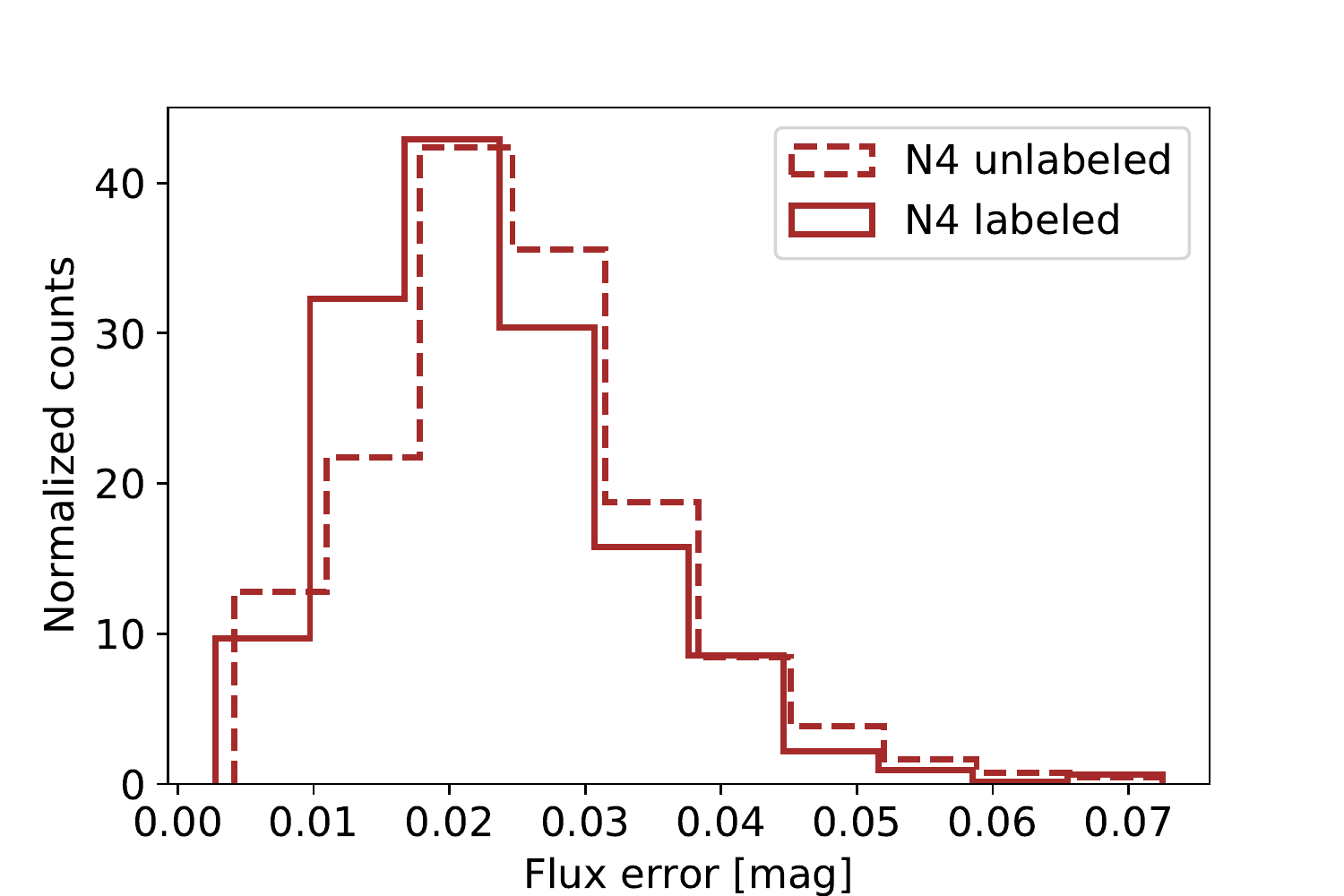}}
\caption{Distribution of the flux measurements and its uncertainties in N4 passband for the $6\_color$ feature set data used for training in Sect.~\ref{main_comparison} and for generalization in Sect.~\ref{best_classifier}.}
\label{fig:fluxN4S7_6col}
\end{figure}

\begin{figure}[h!]
\centering     
\subfigure[S7 and S11 flux distribution of the objects from the generalization and training samples.]{\label{fig:a}\includegraphics[width=80mm]{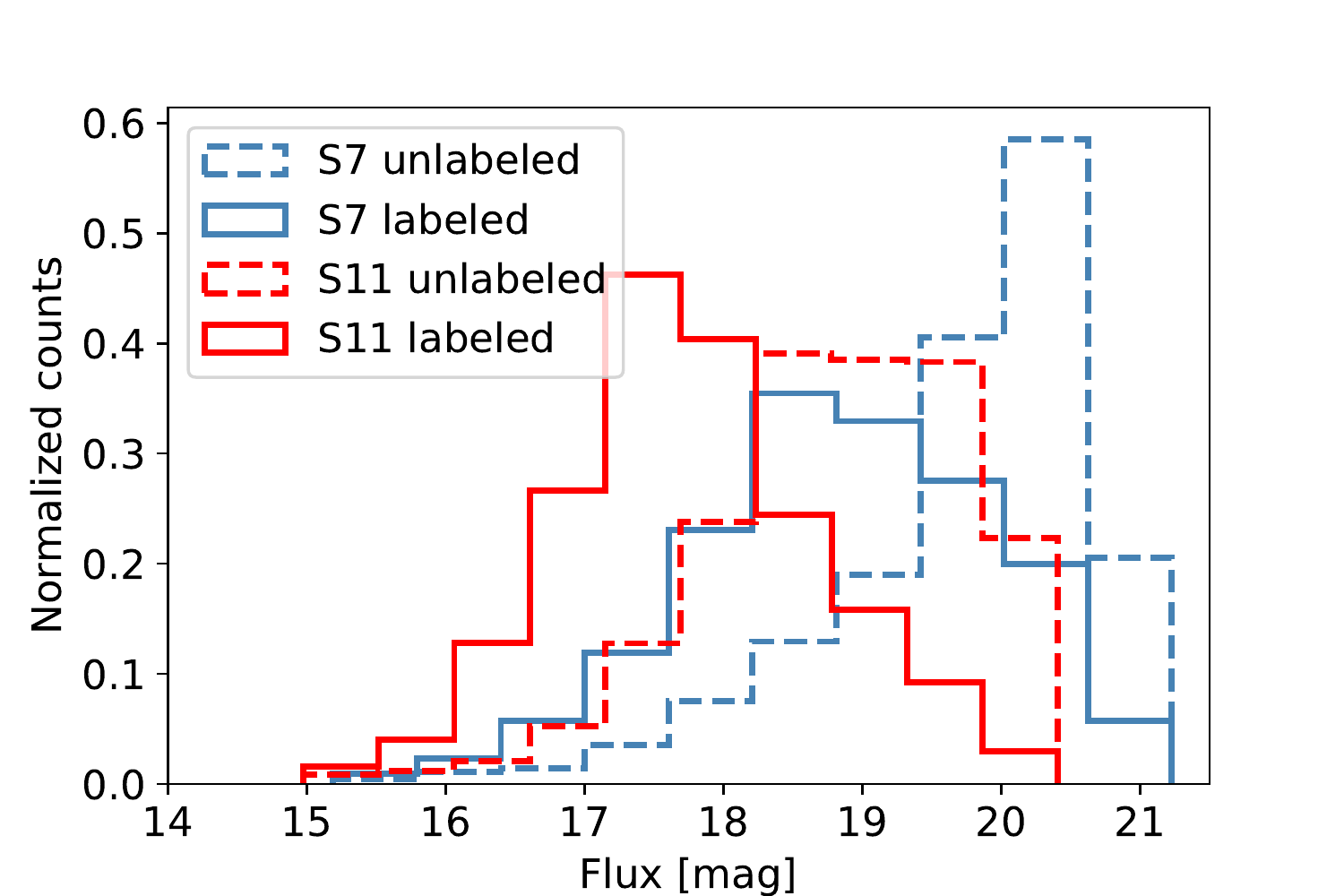}}
\quad
\subfigure[S7 and S11 flux measurement uncertainty distribution for objects from training and generalization samples.]{\label{fig:a}\includegraphics[width=80mm]{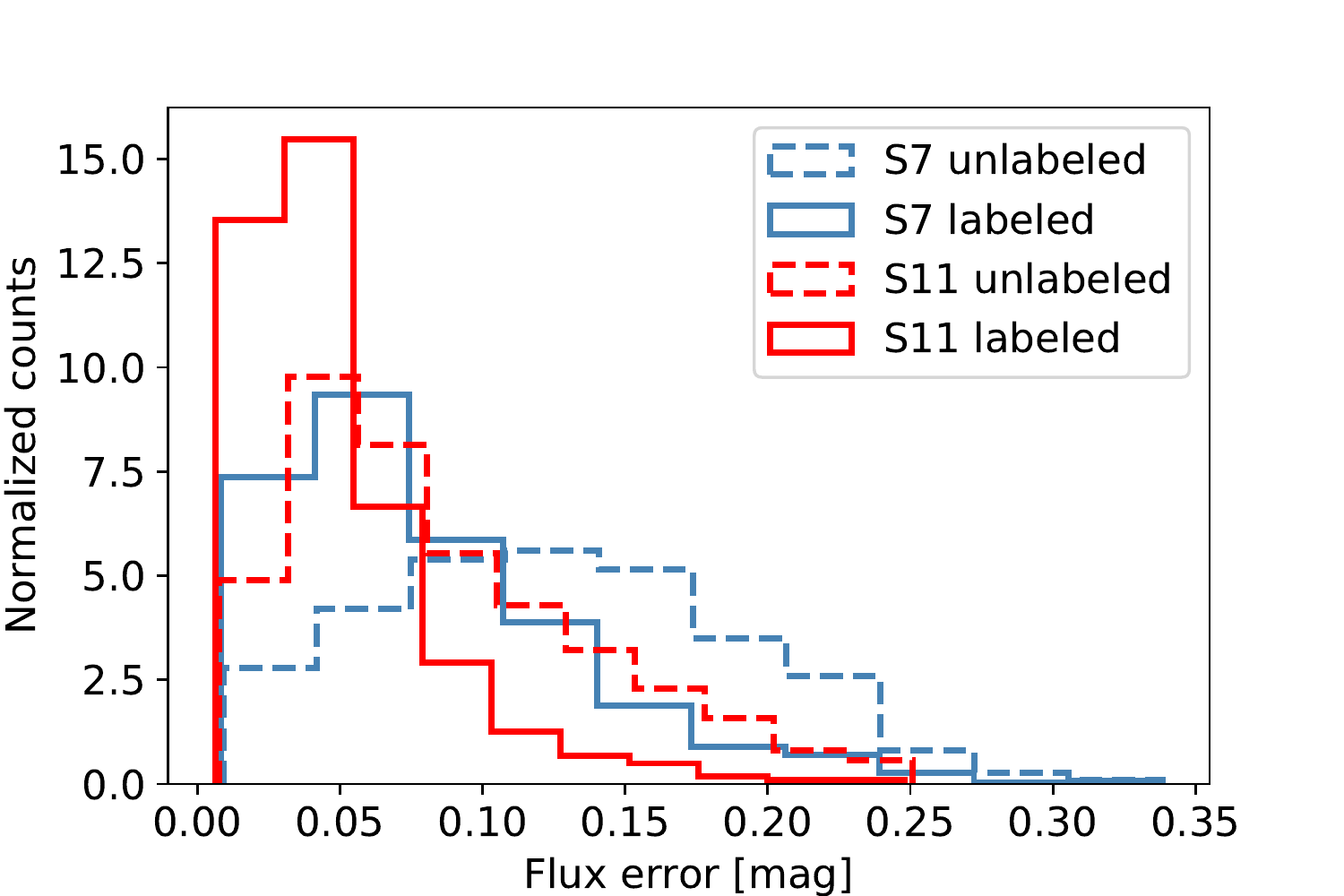}}
\caption{Distribution of the flux measurements and its uncertainties in S7 and S11 passbands for the $6\_color$ feature set data used for training in Sect.~\ref{main_comparison} and for generalization in Sect.~\ref{best_classifier}.}
\label{fig:fluxS11L15_6col}
\end{figure}

\begin{figure}[h!]
\centering     
\subfigure[L15 and L18 flux distribution of the objects from the generalization and training samples.]{\label{fig:a}\includegraphics[width=80mm]{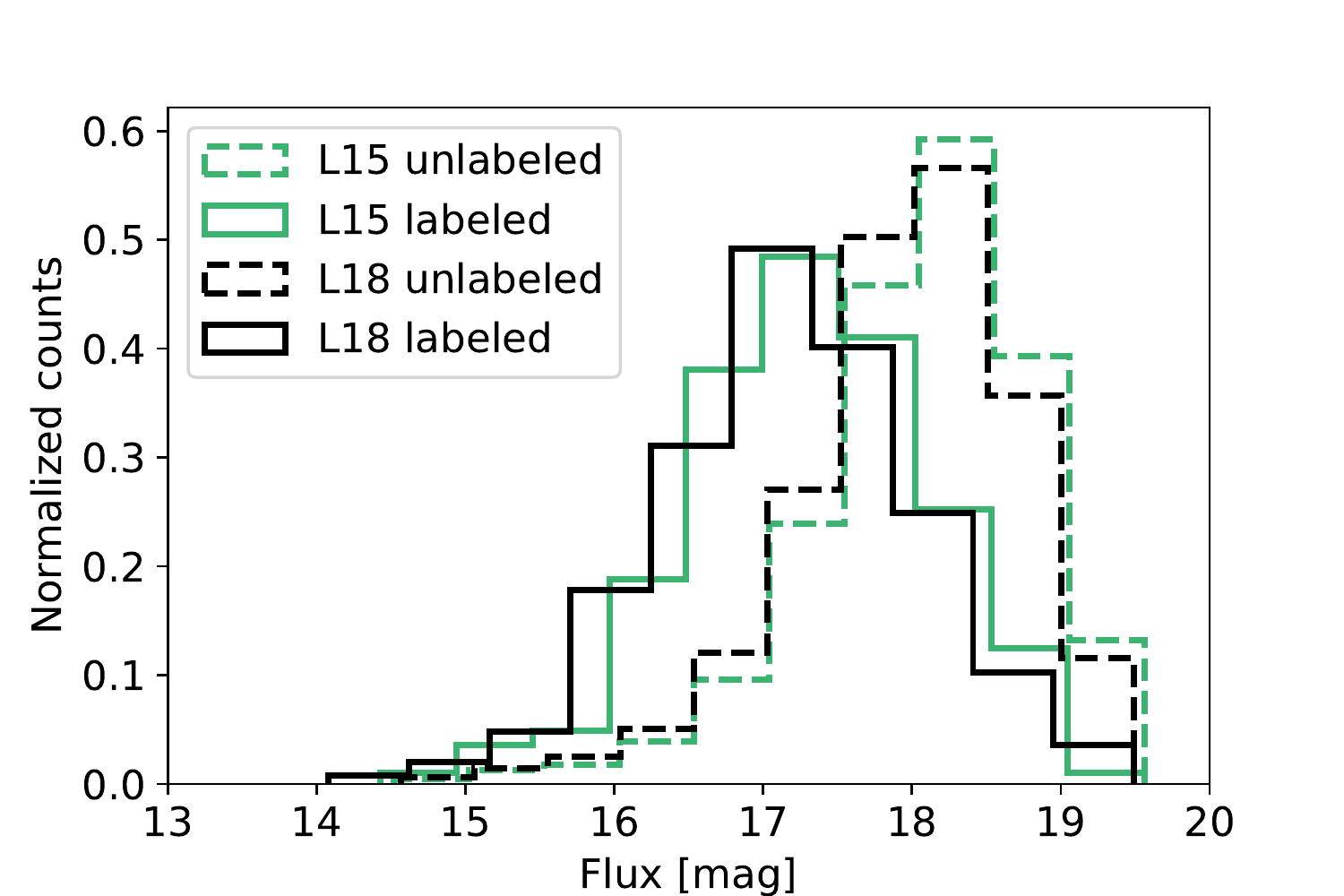}}
\quad
\subfigure[L15 and L18 flux measurement uncertainty distribution for objects from training and generalization samples.]{\label{fig:a}\includegraphics[width=80mm]{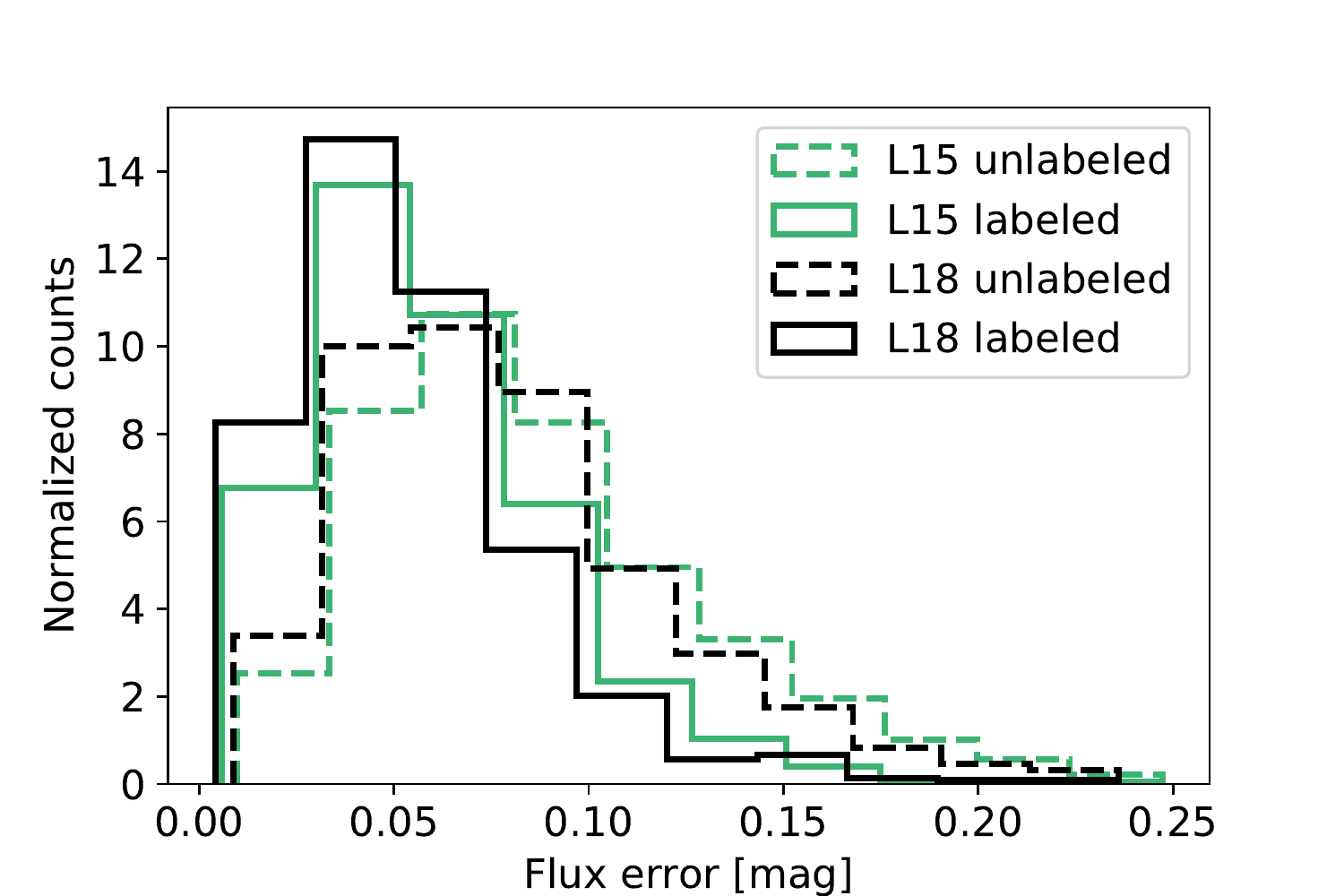}}
\caption{Distribution of the flux measurements and its uncertainties in L15 and L18 passbands for the $6\_color$ feature set data used for training in Sect.~\ref{main_comparison} and for generalization in Sect.~\ref{best_classifier}.}
\label{fig:fluxL18_6col}
\end{figure}

\begin{figure}[h!]
\centering     
\includegraphics[width=80mm]{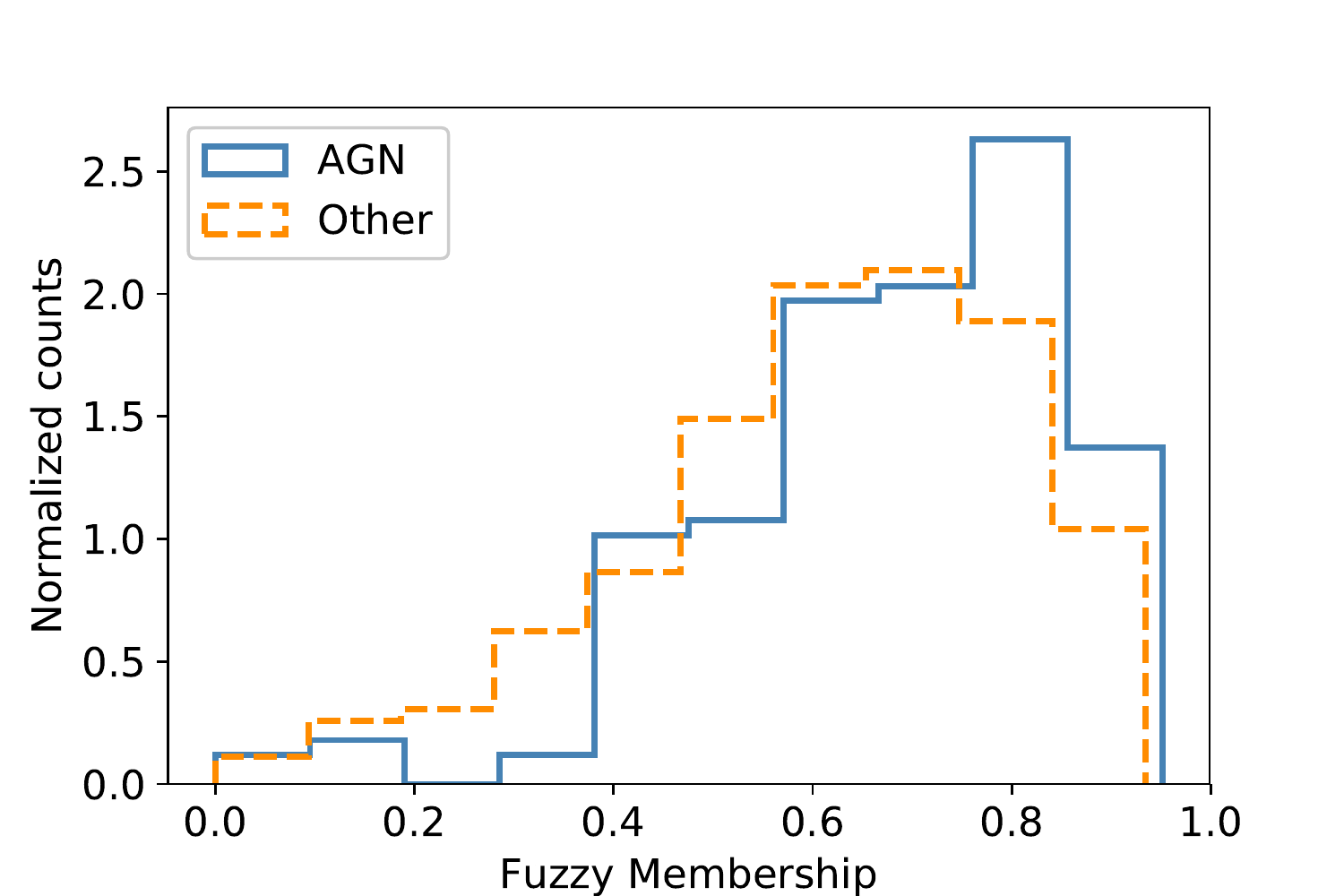}
\caption{Distribution of the fuzzy memberships for AGN and Other classes for the $6\_colors$ feature set used for training in Sect.~\ref{main_comparison}.}
\label{fig:fuz_ap}
\end{figure}

Properties of the data used for the $6\_colors$ feature set. Figures~\ref{fig:fluxN2N3_6col}-~\ref{fig:fluxL18_6col} show flux distribution in particular passbands for labeled and generalization samples as well as distribution of corresponding flux uncertainties. Figure~\ref{fig:fuz_ap} shows distribution of fuzzy memberships for particular classes.

\end{document}